\documentclass[12pt,a4paper]{article}
\usepackage[affil-it]{authblk}
\usepackage{graphicx}
\usepackage[numbers]{natbib}
\usepackage{adjustbox}

\begin{document}

\title{Smallest Fullerene-like Structures of Boron with Cr, Mo, and W Encapsulation} 

\author[1,2]{Amol B. Rahane%
   \thanks{Electronic address:\texttt{amol\_rahane2000@yahoo.com,amolrahane@kthmcollege.ac.in}}}
\author[3]{Pinaki Saha}%
\author[3,4]{N. Sukumar}%
\author[1,4]{Vijay Kumar%
  \thanks{Electronic address: \texttt{kumar@vkf.in, kumar.vkf@gmail.com, vijay.kumar@snu.edu.in}; Corresponding author}}
\affil[1]{Dr. Vijay Kumar Foundation, 1969, Sector 4, Gurgaon - 122 001, Haryana, India, Tel: +91 124 407 9369}
\affil[2]{Department of Physics, K. R. T. Arts, B. H. Commerce and A. M. Science (KTHM) College, Nashik - 422 002, Maharashtra, India.}
\affil[3]{Department of Chemistry, School of Natural Sciences, Shiv Nadar University, NH-91, Tehsil Dadri, Gautam Budhha Nagar 201 314, Uttar Pradesh, India.}
\affil[4]{Center for Informatics, School of Natural Sciences, Shiv Nadar University, NH-91, Tehsil Dadri, Gautam Budhha Nagar 201 314, Uttar Pradesh, India.}

\maketitle

\newpage
\begin{abstract}

Using density functional theory calculations, we study doping of a Cr, Mo, and W atom in boron clusters in the size range of 18-24 atoms and report the finding of metal atom encapsulated fullerene-like cage structures with 20 to 24 boron atoms in contrast to a fullerene-like structure of pure boron with 40 atoms. Our results show that bicapped drum-shaped structures are favoured for neutral Cr@B$_{18}$, Mo@B$_{20}$, and W@B$_{20}$ clusters whereas a drum-shaped structure is preferred for neutral, cation, and anion of Mo@B$_{18}$ and W@B$_{18}$. Further, we find that B$_{20}$ is the smallest cage for Cr encapsulation, while B$_{22}$ is the smallest symmetric cage for Mo and W encapsulation and it is {\it magic}. Symmetric cage structures are also obtained for Mo@B$_{24}$ and W@B$_{24}$. A detailed analysis of the bonding character and molecular orbitals suggests that Cr@B$_{18}$, Cr@B$_{20}$, M@B$_{22}$ (M = Cr, Mo, and W) and M@B$_{24}$ (M = Mo and W) cages are stabilized with 18 $\pi$-bonded valence electrons whereas the drum-shaped M@B$_{18}$ (M = Mo and W) clusters are stabilized by 20 $\pi$-bonded valence electrons. Calculations with PBE0 functional in Gaussian 09 code show that in all cases of neutral clusters there is a large highest occupied molecular orbital-lowest unoccupied molecular orbital (HOMO-LUMO) gap. In some cases the lowest energy isomer of the charged clusters is different from the one for the neutral. We discuss the calculated infrared and Raman spectra for the neutral and cation clusters as well as the electronic structure of the anion clusters. Also we report results for isoelectronic anion and neutral clusters doped with V, Nb, and Ta which are generally similar to those obtained for Mo and W doped clusters. These results would be helpful to confirm the formation of these doped boron clusters experimentally.

\end{abstract}


\section{INTRODUCTION}

The recent finding of fullerene-like empty cage structure\cite{B40} of B$_{40}$ using a 
combined experimental and computational study as well as the occurrence of quasi-planar 
structures for many boron clusters has spurred a lot of interest in 
experimental and theoretical studies of boron clusters to understand their growth behavior and stability. 
Also a layer of boron, called $\alpha$-sheet as well as quasi-planar structures of clusters are stabilized with hexagonal holes in otherwise triangular structures. These results are exciting because the possibility of the formation of carbon-like nanotubular or graphene-like planar structures based on a hexagonal lattice was ruled out due to the deficiency of electrons in boron which is known to favor three center bonding. It has been suggested that the stability of $\alpha$-sheet as well as quasi-planar structures of clusters is due to a mixture of triangular and hexagonal network that is energetically favorable over only a triangular network which has excess of electrons. Several studies on small boron clusters having up to around 36 atoms show that 
they have planar or quasi-planar or tubular structures as their ground state\cite{Zhai2003a,Zhai2003b,Kiran2005,Lau2005,Huang2010,Popov2013,Oger2007} with the exception of B$_{14}$ for which a cage-like structure has been proposed\cite{Cheng2012}. A double-ring tubular (DRT) structure has been suggested for neutral B$_{20}$\cite{Kiran2005} and B$_{24}$\cite{Chacko2003} while a cage structure has been suggested for B$_{38}$\cite{B38} using 
particle swarm algorithm combined with density functional theory (DFT) calculation. A recent study by Chen {\it et al.} has also identified a cage structure with axially chiral feature for B$_{39}$\cite{B39} and another recent theoretical 
study\cite{b28-zhao2015} on neutral B$_{28}$ suggests a filled cage structure to be the lowest in energy while a planar isomer is nearly degenerate. Also, although for neutral B$_{40}$ a fullerene-like cage structure is the most stable one, a planar isomer is favoured for B$_{40}$ anion and therefore a quasi-planar isomer competes in energy for B$_{40}$. A similar situation may arise for other sizes as well that charged clusters have another competing structure. We ask the question if smaller cages of boron can be stabilized such as by metal (M) atom encapsulation. 

Studies on a large number of bulk boron compounds show that an empty center icosahedral cage of B$_{12}$ is the major building block of their structures. However, an isolated B$_{12}$ icosahedral cage is not stable due to the 
availability of 36 valence electrons, while 26 electrons are required by Wade's rules\cite{Wade1971} to stabilize this cage. One way to stabilize the icosahedral cage is to attach ligands such as hydrogen atoms exohedrally. 
As an example, cage-like borane structure B$_{12}$H$_{12}$$^{2-}$ is stabilized by 26 valence electrons excluding those in the B-H bonds.\cite{Lipscomb1969} Another way to form cage structures is to dope endohedrally an M atom that may interact with boron atoms strongly and stabilize 
cage structures. Such a strategy has been successfully used to stabilize non-carbon cage structures such as those of silicon and other elements.\cite{Kumar2001,Kumar2002,Kumar2002-2,Kumar2002-3,Kumar2003-1,Kumar2003-2,Kumar2003-3,Kumar2007} In particular, exceptional stability has been suggested for Zr@Si$_{16}$ 
fullerene and Ti@Si$_{16}$ Frank-Kasper polyhedron structures\cite{Kumar2001} that have been subsequently realized in laboratory and even assemblies have been formed.\cite{nakajima} The size of a B$_{12}$ icosahedron is too small to encapsulate an M atom. As boron atom is smaller 
in size compared to a silicon atom, a possibility to form boron cages may lie around the size of about 20 atoms. A recent independent study has indeed proposed stabilization 
of M doped cages for B$_{24}$ by encapsulation of Mo and W, whereas for Cr a less symmetric configuration has been reported for this size.\cite{Lv2015} Disk-like or wheel-shaped structures have also been predicted for boron clusters 
having up to about 11 boron atoms by doping a transition M  
atom\cite{Romanescu2011,Li2011,Romanescu2012,Galeev2012,Romanescu2013-1,Romanescu2013-2,Zhao2014,Zhao2015} as well as other elements.\cite{saha2016} The stability of some disk-shaped clusters has been correlated with electronic shell closing at 12 valence electrons.\cite{saha2016} Further, recently bowl-shaped\cite{SDLi2006,Boyukata2011,Gu2012,Pham2015,saha2017} and drum-shaped\cite{laisheng2015,saha2017} structures of boron have been predicted to be stabilized by M atom dopants such as M@B$_{14}$ drum with M = Cr, Fe, Co, and Ni, and also with 16 boron atoms such as Co@B$_{16}^-$. 

We have studied M atom encapsulated boron clusters in the size range of 18 to 24 atoms in order to find the smallest cage of free boron clusters besides the drum structures. Note that in this size range, pure boron clusters have quasi-planar or tubular structures. As the bonding in boron is stronger compared with silicon, we considered strongly interacting transition M atoms with half-filled $d$ states such as Cr, Mo, and W to explore the possibility of stabilizing boron cages in this size range. Based on our systematic study, we predict fullerene-like cages Cr@B$_{20}$, M@B$_{22}$ and M@B$_{24}$ with M = Cr, Mo, and W in which one M atom is encapsulated in the cage. Also we find bicapped drum structures for Cr@B$_{18}$ and M@B$_{20}$ (M = Mo and W) and a drum structure for M@B$_{18}$, M = Mo and W. The stability of these structures has been studied by analysing the molecular orbitals (MOs) as well as the electronic charge density. Our results suggest that Mo and W are well suited to produce endohedrally doped novel cage structures of boron as the doping of M atom has the effect of increasing the binding energy significantly. The doping of Cr atom also leads to cage formation, but the binding energy of the doped clusters has similar values as for the pure boron clusters. We also present results for IR and Raman spectra of the neutral and cation clusters as well as for the electronic structure of anion clusters that would help to identify the atomic structure of these doped clusters when experimental data may become available.

\section{COMPUTATIONAL DETAILS}
We used generalized gradient approximation (GGA) of Perdew, Burke, and Ernzerhof (PBE)\cite{PBE} for the exchange-correlation functional and projector augmented wave (PAW) pseudopotential plane wave method\cite{blochl,kresse} in Vienna $\it{ab}$ $\it{initio}$ simulation package (VASP)\cite{vasp} to explore several isomers for the doped boron clusters. The calculations were considered to be converged when the absolute value of the force on each ion was less than 0.005 eV/$\AA$ with a convergence in the total energy of 10$^{-5}$ eV. Further calculations were performed for the lowest energy isomers of the neutral clusters using Gaussian09 code\cite{g09} and PBE0 functional. Also, we have calculated the vibrational modes of cation and neutral clusters using Gaussian09 code, and in almost all cases we obtained real frequencies suggesting the dynamical stability of the obtained atomic structures. For these calculations we used B3PW91 hybrid exchange-correlation functional as well as PBE0 functional and 6-311+G basis set\cite{Wachters1970,Hay1977} for Cr doping and LANL2DZ basis set\cite{Hay1985,Hay1985a} for Mo and W while 6-311+G basis set has been used for B atoms in all the cases. The atomic structures were optimized and used to calculate IR and Raman spectra for the neutral and cation clusters. Calculations using PBE0 exchange-correlation functional on the lowest energy isomers of the pure and M doped boron clusters also suggest stabilization of boron cages with M encapsulation. The bonding characteristics have been studied by performing adaptive natural density partitioning (AdNDP)\cite{AdNDP2008} analysis at the PBE level of the theory and using the same basis sets in the Gaussian09 code and also from the Laplacian L = -(1/4)$\Delta^2 \rho(\bf{r})$, of the electronic charge density $\rho(\bf{r})$ using AIMALL.\cite{AIMALL} We also calculated the electron localization function (ELF) using the charge density distribution obtained from VASP for the lowest energy isomers of the neutrals and the electron localization-delocalization index using AIMLDM script.\cite{AIMLDM} Further calculations have been done on cation and anion clusters using PBE0 functional in Gaussian09 code. Some results are also included for the isoelectronic anions of V, Nb, and Ta doped clusters. The MOs have been analysed using the Gaussian 09 code. We used VESTA 3,\cite{VESTA} Molekel 5.4,\cite{Molekel}, XCrysden 1.5 \cite{XCrysden}, AIMALL, and Gaussview\cite{Gaussview} for visualization. In all cases of charged and neutral clusters the atomic structures were again optimized when using Gaussian09 code.

\section{RESULTS and DISCUSSION}

\subsection{Atomic structures and binding energies}

We studied doping of an M atom (M = Cr, Mo, and W) in several different structures of boron clusters in the size range of B$_{18}$ to B$_{24}$ in the hope of finding the smallest cage of boron. To begin with we doped an Mo atom inside a 16-atom Frank-Kasper polyhedron of boron because all the faces in this structure are triangular and boron has preference to form 3-center bonds as in an icosahedron. But upon optimization it ends up in an open structure indicating that the formation of a boron cage for this size is unlikely with endohedral doping of an Mo atom. However, recently drum-shaped clusters of 16 boron atoms have been obtained with the doping of a Co atom, while bowl-shaped clusters have been shown\cite{saha2017} to be formed for many of the 3d, 4d, and 5d transition M atoms. Also 14-atom drum-shaped boron clusters have been shown\cite{saha2017} to be favored by doping of some 3d transition M atoms such as Cr, Fe, Co, and Ni. These results suggest that a cage structure, besides drum-shaped clusters, is unlikely to be favorable for a 16-atom boron cluster with these M atoms. 

Next we focussed our attention on exploring cage formation with 18 and 20 boron atoms. We added one Mo (and also Cr and W) atom on the planar B$_{18}$ isomer and at the center of the DRT structure of B$_{20}$. Upon optimization, the planar B$_{18}$ structure transforms into a bowl-shaped isomer as shown in Figure S1 (see isomer III and other similar bowl-shaped structures IV and VIII) in Supplementary Information. This clearly showed that the doping of these M atoms tends to form a cage of boron. We rearranged the atoms and tried to construct a cage-like structure of boron but upon re-optimization it becomes again a bowl-shaped open structure indicating that more boron atoms are needed to complete a cage. Some such open as well as cage type structures are shown in Fig. S1 in Supplementary Information, but all of these lie higher in energy than the lowest energy isomer. Figure S1(I) in Supplementary Information and Fig. \ref{fig:B18-B24}(a) show the lowest energy isomer for CrB$_{18}$ which is a bicapped 16-atom drum-shaped structure. On the other hand for MoB$_{18}$ and WB$_{18}$, a tubular drum-shaped structure shown in Fig. \ref{fig:B18-B24}(b) and (c), respectively, and also in Fig. S1(II) in Supplementary Information, has the lowest energy, while the bicapped drum-shaped isomer lies 0.36 eV and 0.42 eV higher in energy for Mo and W, respectively. On the other hand the drum-shaped isomer of Cr@B$_{18}$ lies 1.25 eV higher in energy compared with the lowest energy isomer. In the case of MB$_{20}$ (M = Cr, Mo, and W), we attempted many structures, both open and cage type, and these are shown in Fig. S2 in Supplementary Information. We added two boron atoms in a symmetric fashion to the bowl-shaped MoB$_{18}$ cluster and optimized the atomic structure. The resulting structure was found to be lower in energy than the one obtained from the doping of an Mo atom in B$_{20}$ DRT structure. However, the optimized structure is still open, again suggesting that 20 boron atoms are not sufficient to form a cage with Mo encapsulation. Among the different structures we explored, it is found that a bicapped 18-atom drum-shaped isomer (Fig. \ref{fig:B18-B24}(e) and (f)) has the lowest energy for Mo and W doping. However, as we shall show below, Cr is able to form a cage (Fig. \ref{fig:B18-B24}(d)) with 20 boron atoms, and it is the smallest cage of boron that we have obtained. 

It is clear from the above that there is room to accommodate a few more B atoms on the bicapped drum of Mo@B$_{20}$ to form a cage. Therefore, we added two more B atoms on the other side of the drum so that the two capping dimers form a cross leading to Mo@B$_{22}$ cluster. Also we added four boron atoms to MoB$_{20}$ open structure to form a Mo@B$_{24}$ cage structure in the hope of finding a symmetric isomer, as intuitively a 22-atom cage structure is unlikely to be very symmetric. The optimized structures show that the Mo@B$_{22}$ isomer converges with the same local structure while the optimized atomic structure of Mo@B$_{24}$ had a cage form, but it was slightly lower in symmetry. We changed the positions of two boron atoms in this Mo@B$_{24}$ structure in order to create a symmetric structure and re-optimized it. In two such steps we obtained a very symmetric cage structure of Mo@B$_{24}$ as shown in Figure \ref{fig:B18-B24} (k). This is about 2 eV lower in energy compared with the initially optimized structure (without rearranging boron atoms). The pathway starting from B$_{18}$ to the lowest energy isomer of Mo@B$_{24}$ is shown in Figure \ref{fig:B24-path}. The lowest energy structure for Mo@B$_{24}$ has (D$_{3h}$) symmetry and a large highest occupied molecular orbital-lowest unoccupied molecular orbital (HOMO-LUMO) gap (GGA) of 2.46 eV indicating its high stability. We tried several other isomers for Mo@B$_{24}$ but all of them lie higher in energy than the D$_{3h}$ cage isomer, except in one case where we considered an Mo atom doped at the center of a truncated cube and it converged to the same lowest energy D$_{3h}$ isomer. This gave us further support that our structure may be the lowest in energy. The pathway is indicated in Figure \ref{fig:B20-22-24}. Interestingly the same isomer has been also independently obtained for Mo@B$_{24}$ by Lv et al.\cite{Lv2015} using particle swarm optimization algorithm implemented in the CALYPSO package, which involved calculations of a few thousand structures. In these calculations hybrid PBE0 method has been used for the structure optimization with 611G* basis set in Gaussian 09 package. This gave us further confidence that our structure may indeed be the lowest energy structure for Mo@B$_{24}$. Further calculations on W doping led to the same isomer to be of the lowest energy. The optimized structures of some of the isomers we attempted for M@B$_{22}$ and M@B$_{24}$ (M = Cr, Mo, and W) are shown in Figs. S3 and S4, respectively in Supplementary Information. It can be noted that for Cr doping many isomers lie in a narrow window of energy. 

After obtaining neutral Mo@B$_{24}$ and W@B$_{24}$ doped boron clusters, both of which are almost identical, we asked ourselves the question if we can stabilize a smaller cage of boron, and in particular if a 24 boron-atom cage is the smallest cage for Mo and W. It is to be noted that when Cr is doped in the lowest energy isomer of Mo@B$_{24}$, the optimized structure is a somewhat distorted cage (Fig. \ref{fig:B18-B24} (j)) in which the Cr atom does not interact with all the boron atoms optimally. This is because the size of a Cr atom is not optimal for the B$_{24}$ cage as it is slightly smaller than Mo and W atoms. The natural question we further asked was: what is the smallest cage stabilized by Cr. To address this question we further explored different isomers for the size of 22 boron atoms, as in the 24-atom cage we can remove two B atoms and still keep the cage structure, and also some cage structures for 20 boron atoms. Below we first discuss the formation of a B$_{20}$ cage and then the B$_{22}$ cage.

\begin{figure}
\centering
\includegraphics[width=0.85\linewidth]{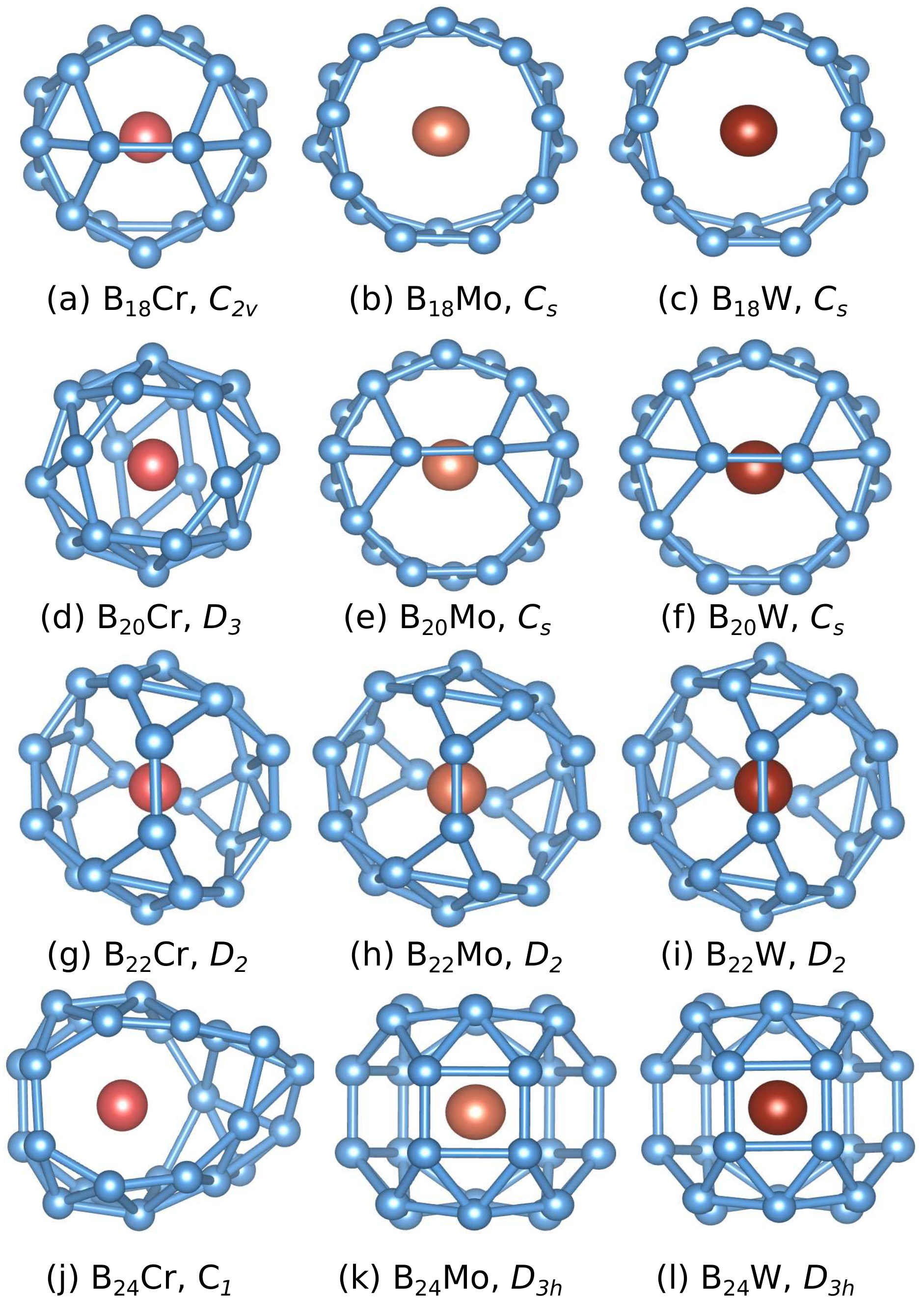}
\caption{Atomic structures of the lowest energy configurations of clusters having 18, 20, 22, and 24 boron atoms doped with a Cr, Mo, and W atom. In each case the symmetry is also given.}
\label{fig:B18-B24} 
\end{figure}

\begin{figure}
\centering
\includegraphics[width=0.9\linewidth]{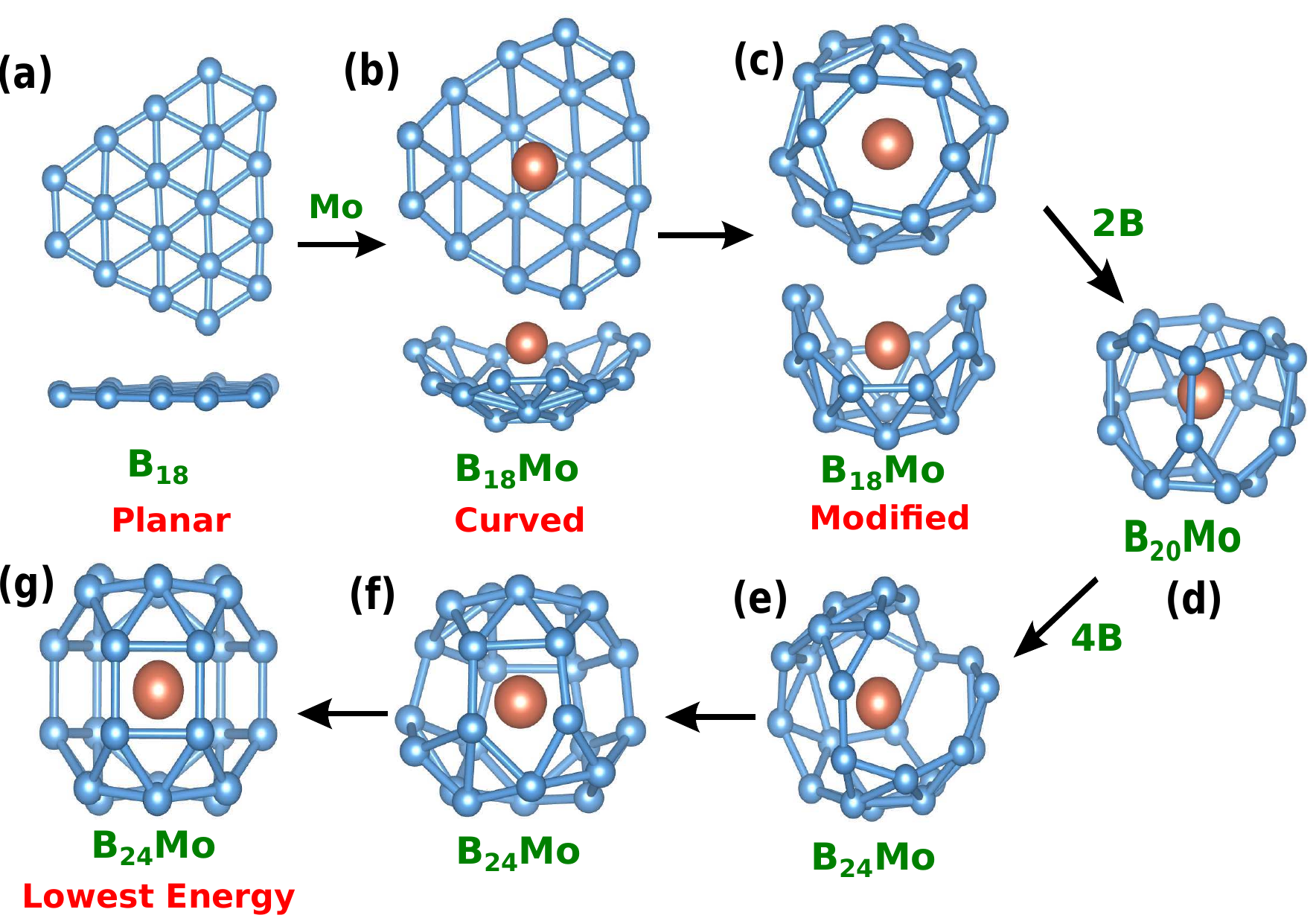}
\caption{Atomic structures showing the pathway from (a) planar B$_{18}$ to (g) D$_{3h}$ Mo@B$_{24}$ cage clusters. (b) is the optimized structure when an Mo atom is placed above the planar isomer (a). (c) is slightly modified structure obtained from (b) by adjusting some atoms in an effort to make a cage. In going from (c) to (d) two boron atoms are added and then four boron atoms are added in the configuration (d). The resulting structure (e) is not symmetric. (f) and (g) are obtained by adjusting B atoms in (e) in order to close the cage and make the cluster symmetric.}
\label{fig:B24-path} 
\end{figure}

\begin{figure}
\centering
\includegraphics[width=0.9\linewidth]{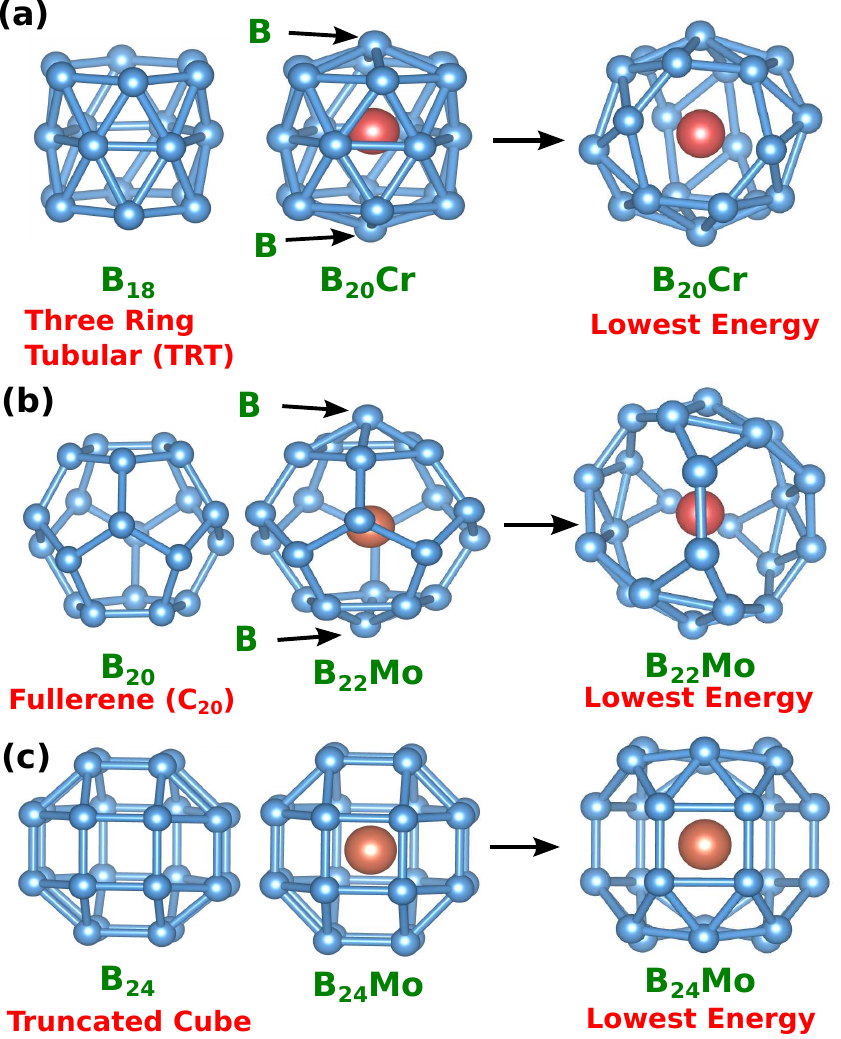}
\caption{Optimized atomic structures for (a) D$_3$ Cr@B$_{20}$, (b) D$_2$ Mo@B$_{22}$, and (c) D$_{3h}$ Mo@B$_{24}$ clusters from their respective starting atomic configurations, namely 18-atom three-ring tubular structure which was capped with two B atoms, a 20-atom dodecahedron which was capped with two B atoms, and 24-atom truncated cube, respectively. In each case Mo atom was placed inside the cage.}  
\label{fig:B20-22-24} 
\end{figure}

In order to study the doping of a Cr, Mo, and W atom in B$_{20}$ cluster, we tried several different initial structures including C$_{20}$ type B$_{20}$ fullerene with all pentagonal faces, quasi-planar, and tubular structures. All of them were optimized and also atoms were rearranged to possibly find symmetric lowest energy structures for all the three M atom dopants. The optimized structures are shown in Fig. S2 in Supplementary Information. First note that for Cr, the CrB$_{18}$ structure remains open and forms a bicapped-drum structure as discussed above, suggesting that eighteen atoms of boron are not sufficient to form a cage. The lowest energy isomers for MB$_{18}$, (M = Cr, Mo, and W) show (Figure \ref{fig:B18-B24}(a)-(c)) that the structure of CrB$_{18}$ is different from the one for the MoB$_{18}$ and WB$_{18}$ clusters. CrB$_{18}$ structure has two boron atoms capped on to a B$_{16}$ DRT structure forming a bicapped-drum structure with $C_{2v}$ symmetry. This structure has two pentagons arranged in base sharing fashion. For Cr@B$_{20}$, we obtained a $D_3$ symmetric cage structure as shown in Figure \ref{fig:B18-B24}(d) to be of the lowest energy. This smallest cage in our calculations has three empty boron hexagons that are inter-linked through three 2-atom chains to two capped boron hexagons. However, for MoB$_{20}$ and WB$_{20}$, a bicapped drum structure with two boron atoms capping an 18-atom boron DRT with $C_s$ symmetry and M atom at the centre has the lowest energy. This structure has one pentagon and one hexagon arranged in base sharing fashion. For the CrB$_{20}$ case, the bicapped drum isomer with $C_s$ symmetry is 0.32 eV higher in energy than the cage structure. Also the $D_3$ fullerene structure for MoB$_{20}$ and WB$_{20}$ cases is, respectively, 1.79 eV and 2.06 eV higher in energy than the lowest energy bicapped drum ($C_s$) structure. We also obtained an unsymmetric cage ($C_1$) structure for Mo@B$_{20}$ and W@B$_{20}$ (see isomer X in Fig. S2 in Supplementary Information) which is 0.88 eV and 1.02 eV, respectively, higher in energy than the lowest energy structure. Several other isomers that we tried are shown in Fig. S2 in Supplementary Information using PBE in VASP. These results suggest that for Mo and W, the cage need to have more than 20 boron atoms.

In order to explore a cage of B$_{22}$, we tried several configurations of neutral Mo@B$_{22}$. The lowest energy configuration has $D_2$ symmetry as shown in Figure \ref{fig:B18-B24}(h). This is obtained by adding two boron atoms in the $C_1$ isomer of Mo@B$_{20}$ with slight rearrangement. This structure has four heptagons that are interlinked through two 2-atom chains and three boron atoms. Another calculation starting from C$_{20}$ fullerene for B$_{20}$ and capping two opposite pentagons with two boron atoms converged to the same $D_2$ structure after optimization for Mo@B$_{22}$. This is shown in Fig. \ref{fig:B20-22-24}(b). We also tried several isomers for Cr and W doping and obtained the same $D_2$ structure for W@B$_{22}$ to be of the lowest energy as shown in Figs. \ref{fig:B18-B24}(i), but for Cr@B$_{22}$, while in VASP calculation, the same isomer is of the lowest energy, a slightly different isomer shown in Fig.\ref{fig:B18-B24}(g) is obtained from Gaussian calculation as the lowest energy one. This structure also shows that less than 22 boron atoms are better to have an optimal cage with Cr doping. Interestingly when this isomer is reoptimized in VASP, it reverts to the same structure as before in VASP. Several isomers including the lowest energy isomers for Cr, Mo, and W doped B$_{22}$ are shown in Fig. S3 in Supplementary Information. As can be seen, there are a few isomers for Cr doping that lie close in energy within about 0.30 eV of each other. The isomer in which the M atom interacts with a quasi-planar isomer of B$_{22}$ (Fig. S3 XVI in Supplementary Information) lies more than 2 eV higher in energy than the lowest energy isomer.

For the case of the B$_{24}$ cage, we further explored different structures for M@B$_{24}$ with M = Cr, Mo, and W. These are shown in Fig. S4 in Supplementary Information. The lowest energy configuration for Cr@B$_{24}$ has C$_1$ symmetry as shown in Fig. \ref{fig:B18-B24}(j). The isomer with D$_{3h}$ configuration for Cr lies 0.25 eV higher in energy. Although our result of the lowest energy structure of Cr@B$_{24}$ is in agreement with that reported by Lv {\it et al.}\cite{Lv2015}, they reported this structure to have quintet spin-multiplicity. But in our calculations a singlet configuration has the lowest energy. The triplet and quintet configurations are, respectively, 0.97 eV and 2.02 eV higher in energy. As stated earlier, the lowest energy isomers for Mo@B$_{24}$ and W@B$_{24}$ are similar and are shown in Figs. \ref{fig:B18-B24}(k) and (l), respectively.

\begin{figure}
\centering
\includegraphics[width=0.8\linewidth]{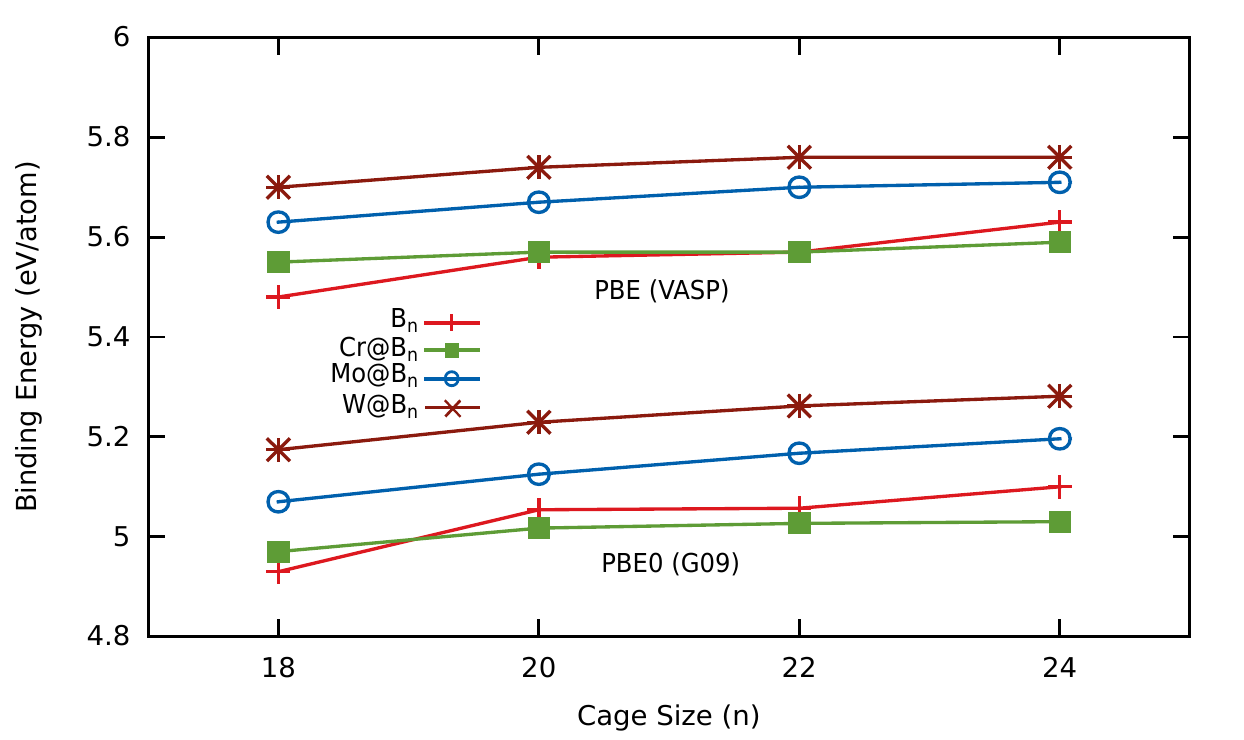}
\caption{Calculated binding energies for the lowest energy isomers.}
\label{fig:BE} 
\end{figure}

It is to be noted that neutral pure boron clusters, B$_n$ with $n$ = 18-24, prefer quasi-planar or tubular structures in their ground states and these become drum-shaped or form cages after doping of a transition M atom. We calculated the energy of the drums/cages by removing the M atom and keeping the atomic positions of the boron atoms the same as in the optimized doped cluster. 
It is seen that the empty cage structures of B$_n$ ($n$ = 20, 22, and 24) lie 1.50-3.50 eV higher in energy than their respective ground state isomers within PBE. Therefore, the interaction of the M atom due to endohedral doping lowers the energy of the cage significantly. The doping energy has been calculated from $\Delta$$E$ = $E(M atom) + E(B_n cage) - E(M@B_n)$ and the values are given in Table 1. In general large values ($>$ 8 eV) of $\Delta$$E$ are obtained for W doping and for Mo in B$_{22}$ cage. 
This also leads to significant enhancement in the binding energy of the doped boron clusters compared with that of the undoped clusters. The binding energy is defined as the energy per atom of the cluster with respect to the energy of the free atoms, [$nE(B) + E(M) - E(M@B_n)]/(n+1)$, and it is shown in Fig. \ref{fig:BE} using PBE in VASP and PBE0 in Gaussian 09 program for the lowest energy isomers of all the sizes and with different dopants. Here $E(B)$, $E(M)$, and $E(M@B_n)$ are the energies of B atom, M atom, and the doped cluster, respectively. The overall trend of the binding energy in both the methods is similar, but the values are slightly lower when PBE0 is used. These results also show that the binding energy increases in going from Cr doping to Mo and then to W, indicating that W doping is the strongest among the three dopants. There is a slight increase in the binding energy in going from 18-atom to 20-atom boron clusters for all the dopants. For Cr doping the binding energy remains nearly the same in going from Cr@B$_{20}$ to Cr@B$_{22}$. However, it slightly increases to 5.59 eV for Cr@B$_{24}$ cluster within PBE. In the case of Mo and W doping the trend shows that the increase in the binding energy in going from Mo@B$_{20}$ to Mo@B$_{22}$ is more than the increase in going from Mo@B$_{22}$ to Mo@B$_{24}$. Similarly, in the case of W doping the binding energy increases in going from W@B$_{20}$ to W@B$_{22}$, but it remains nearly the same for W@B$_{22}$ and W@B$_{24}$ within PBE. Our results suggest that both 20 and 22 boron atom clusters doped with these M atoms are likely to be abundant, as the second derivative of energy ($\Delta$$E$ = 2$E$(M@B$_n$) - $E$(M@B$_{n-2}$) - $E$(M@B$_{n+2}$))/4 is negative for both these sizes. We have also plotted the values of the binding energy for the pure boron clusters with 18, 20, 22, and 24 atoms. These results show that the doping of Mo and W atoms leads to a significant increase in the binding energy of the boron clusters and therefore, Mo and W are favorable to form endohedral cage structures of boron. However, in the case of Cr doping, the binding energy is either nearly the same or marginally higher (in the case of CrB$_{18}$) or lower (Cr@B$_{24}$) than the value for the corresponding pure boron cluster within PBE in VASP (see Table 1). But using PBE0 in Gaussian 09 program the binding energy of CrB$_{n}$ clusters is slightly lower than the values for the pure boron clusters except for the Cr@B$_{18}$ case as shown in Fig. \ref{fig:BE}. Also we find that within PBE0, there is a large HOMO-LUMO gap for Cr@B$_{18}$. Therefore, we conclude that the stabilization of boron cages with Cr doping is less than those of Mo and W doped clusters, and the magnetic moments on the M atom is quenched. Using PBE0 the second derivative of energy ($\Delta E$) for Cr@B$_{20}$ is -0.165 eV, while for MoB$_{20}$ and WB$_{20}$ doping it is -0.018 eV and -0.072 eV, respectively. 
So M@B$_{20}$ is weakly magic for Mo and intermediate for W doping. Also for M@B$_{22}$, it is -0.015 eV, -0.043 eV, -0.050 eV, respectively for M = Cr, Mo, and W. Accordingly Mo@B$_{22}$ and W@B$_{22}$ are likely to be more abundant than Cr@B$_{22}$. Therefore, MoB$_{22}$ is magic within PBE0 while W@B$_{20}$ is likely to be more abundant than W@B$_{22}$. In the following we discuss Cr as well as Mo and W doping to understand the cage stability and bonding characteristics in the doped clusters.

\begin{table}
\begin{center}
\caption{Calculated embedding energy (E$_n$(M)) of the M atom in n-atom boron cage, BE, HOMO-LUMO gap (E$_g$), vertical electron affinity (VEA) and vertical ionization potential (VIP) for the lowest energy isomers of M@B$_n$ (n = 18, 20, 22, and 24; M = Cr, Mo, W) clusters. VEA and VIP are calculated using Gaussian 09 program. For comparison, the binding energy and HOMO-LUMO gap calculated within PBE0 are also given. Results have also been given for the lowest energy isomers of pure B$_n$ clusters ($n$ = 18, 20, 22, and 24).}
\begin{tabular}{|c|ccc|ccccc|}
\hline
&\multicolumn{3}{c|}{PBE} &\multicolumn{5}{c|}{PBE0} \\
\hline
Cluster & E$_n$(M) &BE per atom&E$_g$ &  E$_n$(M) & BE per atom&E$_g$ &VEA&VIP\\
        & (eV)&(eV)&(eV)& (eV)&(eV)& (eV)&(eV)&(eV) \\
\hline
B$_{18}$& - & 5.48 & 0.47 & -&4.93&1.39&-&-\\
B$_{20}$& - & 5.56 & 1.36 & -&5.05&2.72&-&-\\
B$_{22}$& - & 5.57 & 0.29 & -&5.06&1.35&-&-\\
B$_{24}$& - & 5.63 & 1.30 & -&5.10&2.18&-&-\\

Cr@B$_{18}$& 6.67 & 5.55 & 1.24 & 5.65 & 4.97 &2.73  &2.72 &7.63 \\
Mo@B$_{18}$& 8.28 & 5.63 & 0.40 & 7.53& 5.07 & 1.71 &3.33  &7.37\\
W@B$_{18}$& 9.61 & 5.70 & 0.47  & 9.51& 5.17 & 1.71 & 3.37 &7.44\\
Cr@B$_{20}$& 5.62 & 5.57 & 2.14 &  4.28& 5.02 & 3.85 &1.88 & 7.94\\
Mo@B$_{20}$& 7.76 & 5.67 &1.76 & 6.54& 5.12 & 3.21&2.06 &6.99\\
W@B$_{20}$& 9.22 & 5.74 &1.79  & 8.72& 5.23 & 3.31 &2.00 &7.06 \\
Cr@B$_{22}$& 5.29 & 5.57 & 2.26 &  3.99& 5.03 & 3.78 &2.05 & 7.05\\
Mo@B$_{22}$& 8.25 & 5.70 & 2.51 & 7.59& 5.17 & 3.87&1.97&8.01\\
W@B$_{22}$& 9.67 & 5.76 & 2.50 &  9.76& 5.26 & 3.85&1.99&8.03\\
Cr@B$_{24}$& 4.59 & 5.59 & 0.89 &  3.32& 5.03 & 2.35 &3.44&7.11\\
Mo@B$_{24}$& 7.33 & 5.71 & 2.46 &  7.42& 5.20 & 4.40 &1.59&7.86\\
W@B$_{24}$& 8.69 & 5.76 & 2.45 &  9.55& 5.28 & 4.37&1.56&7.89\\
\hline
\end{tabular}
\end{center}
\end{table}

From Table 1 the calculated HOMO-LUMO gaps (E$_g$) within PBE are 1.24 eV, 0.40 eV, and 0.47 eV for MB$_{18}$, M = Cr, Mo, and W, respectively. Note that GGA underestimates E$_g$ and the values obtained using PBE0 are much larger. 
The values of E$_g$ for these cases are, respectively, 2.72 eV, 1.71 eV and 1.71 eV when PBE0 functional is used. Here note the large value for the case of Cr doping. For MB$_{20}$, the HOMO-LUMO gaps increase to 2.14 eV, 1.76 eV, and 1.79 eV, respectively with M = Cr, Mo, and W, and further to 2.26 eV, 2.51 eV, and 2.50 eV for MB$_{22}$ within PBE. These large values of E$_g$ suggest the stability of the cage structures. However, for Cr@B$_{24}$ the HOMO-LUMO gap decreases to 0.89 eV as the cage becomes less symmetric and Cr is not bonded well with all the B atoms. Mo@B$_{24}$ and W@B$_{24}$ clusters also have large E$_g$ (2.46 eV and 2.45 eV, respectively), but the values are slightly lower than for Mo@B$_{22}$ and W@B$_{22}$ cages. The calculated E$_g$ for the M doped B$_{22}$ and B$_{24}$ cage structures using PBE0 are again much larger than the values obtained within PBE, and these are given in Table 1. The largest value (4.40 eV) of E$_g$ within PBE0 is obtained for Mo@B$_{24}$ and a quite similar value for W@B$_{24}$. The largest value (5.28 eV/atom) of the binding energy within PBE0 is obtained for W@B$_{24}$ and a slightly smaller value (5.26 eV/atom) for W@B$_{22}$. This is in contrast to the values of 5.06 eV/atom and 5.10 eV/atom for the binding energies of elemental boron clusters B$_{22}$ and B$_{24}$, respectively, within PBE0. Therefore, M doping enhances the stability of boron clusters and leads to the formation of the cage structures of boron for smaller sizes compared with pure boron clusters.   

\subsection{Bonding characteristics}

In order to understand the bonding character in these structures, we noted that the shortest B-B bond lengths for Cr@B$_{18}$, Mo@B$_{18}$, and W@B$_{18}$ are 1.61 {\AA}, 1.58 {\AA}, and 1.58 {\AA} respectively, whereas B-Cr, B-Mo, and B-W bond lengths are in the range of 2.12-2.36 {\AA}, 2.32-2.63 {\AA}, and 2.33-2.60 {\AA}, respectively using PBE. The shortest B-B bonds are the two-center (2c) sigma bonds. As the cluster size increases, we find that for Cr@B$_{20}$ the B-B and B-Cr bond distances lie in the range of 1.60-1.83 and 2.08-2.30 {\AA}, respectively while for Mo@B$_{20}$ and W@B$_{20}$ the shortest B-B bond length is 1.57 {\AA}. The B-Mo and B-W bond lengths are in the range of 2.13-2.57 {\AA} and 2.14-2.58 {\AA}, respectively. These Mo-B and W-B bond lengths are shorter than for Mo@B$_{18}$ and W@B$_{18}$, respectively. The shorter bonds are formed by the capping boron atoms with the M atom. For the M doped B$_{22}$ cage, some of the B-B bonds are shorter than those in the M doped B$_{20}$ cage. There are two shortest B-B bonds for the B$_{22}$ cage with bond length 1.56 {\AA} for Cr encapsulation, and 1.57 {\AA} for Mo and W encapsulation, but the M-boron bond lengths are slightly elongated and lie in the range of 2.31-2.54 {\AA} for Cr encapsulation and 2.38-2.59 {\AA} for Mo and W encapsulation as compared to those in the respective doped B$_{20}$ cages. In the B$_{24}$ cage the shortest B-B bond length increases to 1.64 {\AA} for Mo and W encapsulation. Also some of the M-boron bond lengths are longer and lie in the range of 2.43-2.55 {\AA} for Mo encapsulation and 2.44-2.55 {\AA} for W encapsulation. For Cr encapsulation in B$_{24}$, the B-B bond lengths are in the range of 1.54-1.84 {\AA} while B-Cr bond lengths have wide variation and are in the range of 2.18-2.66 {\AA}. This is becuase the Cr atom is not at the center of the cage. 

We further calculated the total charge density isosurfaces, ELF, and molecular orbitals for the doped clusters. We also performed AdNDP analysis on some of the clusters. Details of these results are presented in Supplementary Information and in Figs. \ref{fig:MB20-22-24-elf} and \ref{fig:B20Cr-bonding}. Earlier studies on pure boron clusters have suggested the existence of multi-center bonds namely 2c, 3c, 4c, 6c, and 7c bonds using AdNDP analysis\cite{AdNDP2008}. However, in our experience the AdNDP analysis is not straightforward as there is not a unique way to identify multi-center bonds. We were greatly assisted by the analysis of the charge density, molecular orbitals, and ELF. In an earlier study\cite{B84} on B$_{84}$ cluster, the total charge density and ELF were used as tools to understand 2c and 3c bonds in boron clusters along with the AdNDP analysis. This was helpful in further calculating the multi-center bonds with AdNDP. We followed a similar strategy for some of the M doped boron clusters studied here. Further, we have calculated Laplacian of the electron density and bond- as well as ring-critical points to analyse the bonding character.  

\begin{figure}
\centering
\includegraphics[width=0.9\linewidth]{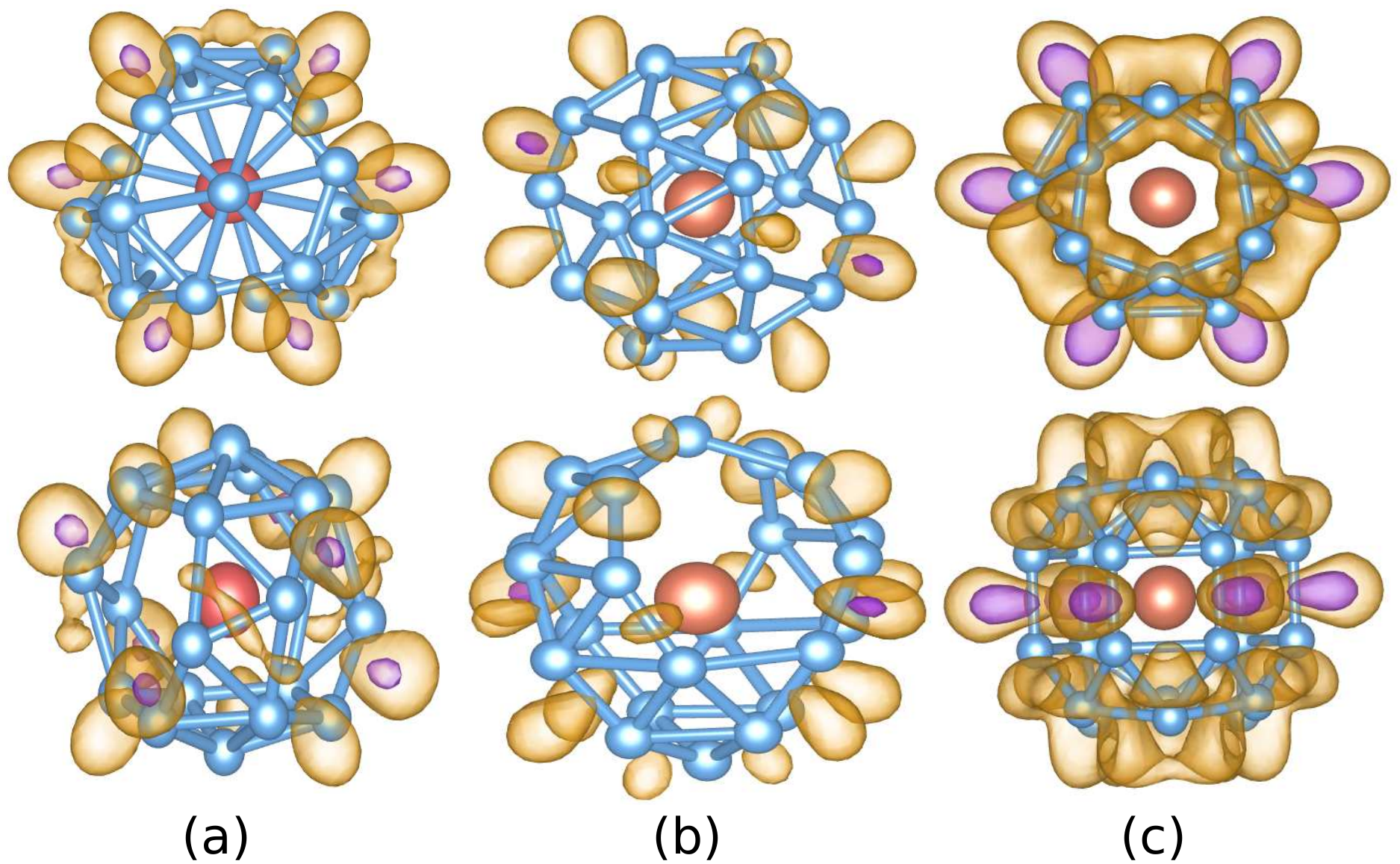}
\caption{Isosurfaces of ELF for (a) Cr@B$_{20}$, (b) Mo@B$_{22}$, and (c) Mo@B$_{24}$. Two orientations are shown in each case. In (a) the lobes shown by purple color have strong localization of charge with high value of ELF (0.94) representing 2c-2e bonds. As the value of the iso-surface is decreased (0.84), these lobes become bigger and new lobes appear. These other lobes shown with yellowish color are 3c or higher center $\sigma$ bonds. In (b) there are 2 purple color lobes with ELF = 0.96 (2c) and sixteen yellowish color lobes with ELF = 0.88 (3c). In (c) there are six 2c (purple color) and eighteen 3c (yellowish) bonds.}
\label{fig:MB20-22-24-elf} 
\end{figure}

Figure \ref{fig:MB20-22-24-elf} shows ELF for the smallest cage Cr@B$_{20}$ and also for Mo@B$_{22}$ and Mo@B$_{24}$ clusters. For Cr@B$_{20}$ the bond length analysis showed that there are six short B-B bonds with 1.60 {\AA} length. These bonds are along the three 2-atom chains joining the three empty heptagons and can be detected in the charge density and ELF plots at a high value of the isosurface (see Fig. S5 in Supplementary Information). A high value of ELF shows strong localization of charge as in covalent bonds. These six lobes are indicated by purple color in Fig. \ref{fig:MB20-22-24-elf}(a) at the ELF value of 0.94. We identify these bonds as six 2c-2e bonds. Further decreasing the ELF value to 0.84, these six lobes become bigger (shown by yellowish color around the purple color lobes) and twelve more lobes can be seen. These twelve additional lobes (yellowish in color in Fig. \ref{fig:MB20-22-24-elf}(a)) represent twelve 3c $\sigma$ bonds. This identification was also supported by the AdNDP analysis as we shall discuss below. Figs. \ref{fig:MB20-22-24-elf}(b) and (c) show ELF plots for Mo@B$_{22}$ and Mo@B$_{24}$, respectively. For Mo@B$_{22}$, the two lobes shown by purple color are obtained at the ELF value of 0.96 and they represent two 2c bonds with the bond length of 1.57 {\AA}. There are sixteen more lobes shown by yellowish color at the ELF value of 0.88. Similar to the case of the B$_{20}$ cage, they represent sixteen 3c bonds. For Mo@B$_{24}$ the analysis of the charge density and ELF shows that this cluster has six 2c $\sigma$-bonds (purple color lobes in ELF) and eighteen 3c $\sigma$-bonds. Further analysis using AdNDP shows six 5c $\sigma$-bonds (see Fig. \ref{fig:MB20-22-24-elf}(c)). The remaining eighteen valence electrons are involved in $\pi$-bonding as well as bonding with the M atom, and can be detected in the AdNDP analysis\cite{Lv2015}. 

\begin{figure*}
\centering
\includegraphics[width=0.4\linewidth]{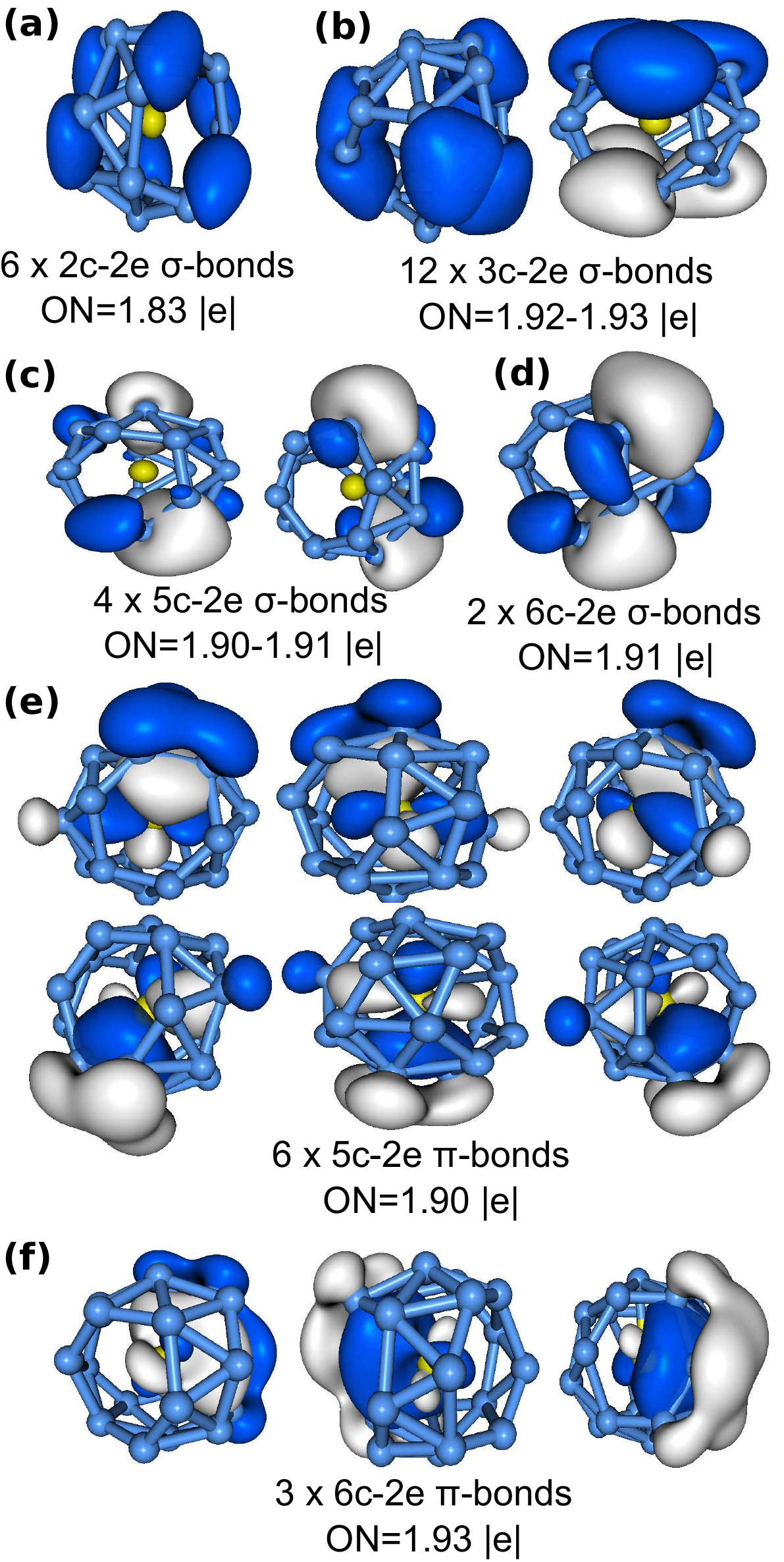}
\caption{Different multi-center bonds in the lowest energy isomer for Cr@B$_{20}$ as obtained from the AdNDP analysis.}
\label{fig:B20Cr-bonding} 
\end{figure*}

We consider Cr@B$_{20}$ cage cluster as the smallest cage, and its AdNDP analysis shows (Fig. \ref{fig:B20Cr-bonding} (a)) that there are six 2c-2e bonds, as also concluded from ELF. Further, there are twelve 3c bonds, six are along the three 2-atom chains, and the remaining six are placed alternatively on the three edges of each capped hexagon as shown in Figure \ref{fig:B20Cr-bonding}(b). These bonds were also detected in the total charge density and ELF analysis as discussed above. In addition to these eighteen $\sigma$-bonds there are four 5c-2e $\sigma$-bonds and two 6c-2e $\sigma$-bonds as shown in Figure \ref{fig:B20Cr-bonding}(c) and (d), respectively. These bonds connect the atoms on the hexagons to the atoms on the three 2-atoms chains. All these 24 $\sigma$-bonds involve bonding among boron atoms. There are six 5c-2e $\pi$-bonds, each connecting three B atoms on one capped hexagon to one B atom on the edge through the Cr atom. There are also three 6c-2e $\pi$-bonds along the three 2-atom chains, each involving six B atoms on the chain and the Cr atom. It is interesting to see that there are altogether nine $\pi$-bonds which involve bonding of the boron atoms on the cage and the Cr atom. This indicates that the stability of the cage due to the Cr atom doping is governed by the completion of an electronic shell with 18 valence electrons, as we shall further show in the following from the analysis of the MOs of Cr@B$_{20}$.

\begin{figure*}
\centering
\includegraphics[width=0.6\linewidth]{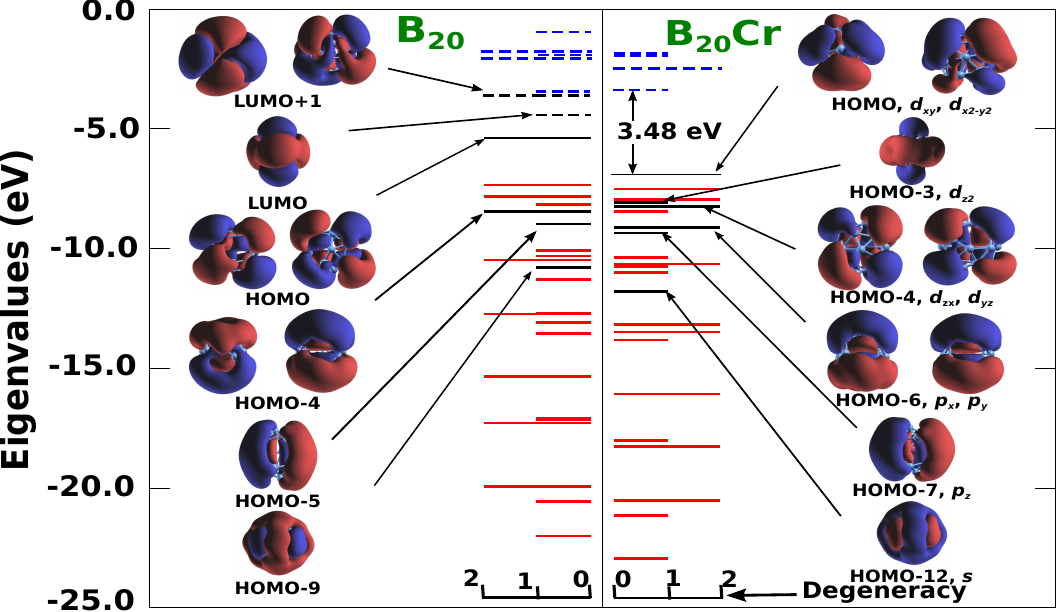}
\caption{The $\pi$-bonded MOs for the bare B$_{20}$ cage and Cr@B$_{20}$ calculated with Gaussian 09 program.}
\label{fig:B20Cr-MO} 
\end{figure*}

\begin{figure*}
\centering
\includegraphics[width=0.6\linewidth]{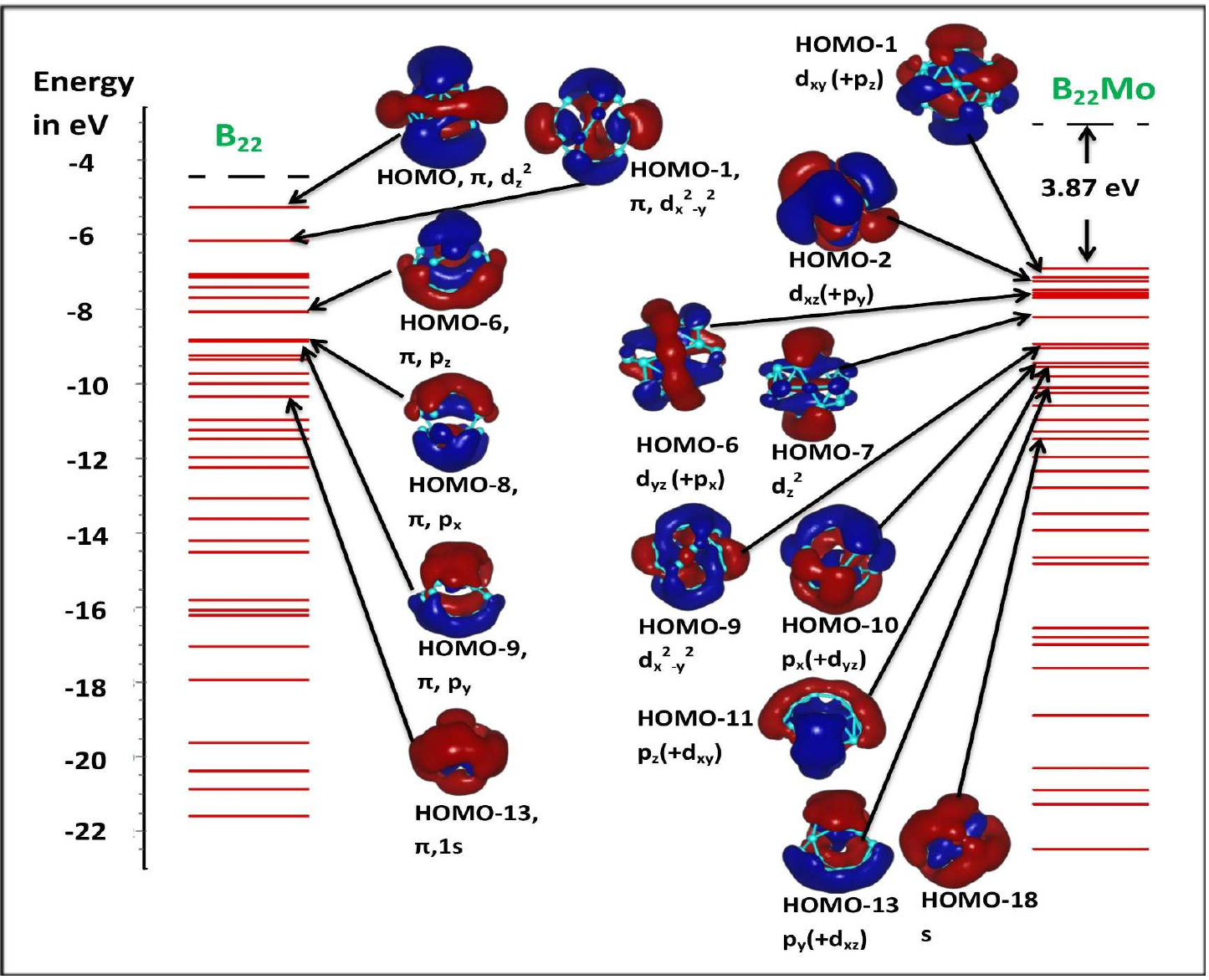}
\caption{Some of the $\pi$-bonded MOs for the bare B$_{22}$ and Mo@B$_{22}$.}
\label{fig:B22-MO} 
\end{figure*}

In order to understand the electronic origin of the stability of these cages, we calculated the MOs for the M doped clusters as well as the corresponding boron cage by removing the M atom and keeping the atomic positions fixed. For the Cr@B$_{20}$ case, six electrons of the Cr atom and twelve $\pi$-bonded electrons from the B$_{20}$ cage contribute to the stability with electronic shell closing at eighteen $\pi$ bonded valence electrons. Figure 7 shows that the bare B$_{20}$ cage has six occupied $\pi$-bonded MOs (doubly degeberate HOMO and HOMO-4, HOMO-5, and HOMO-9). The two HOMO levels have {\it d} angular momentum character. There are three unoccupied $\pi$-bonded MOs (LUMO and LUMO+1) of {\it d} angular momentum character. For Cr@B$_{20}$ these five MOs of {\it d} angular momentum character hybridize with the {\it d} orbitals of Cr atom forming bonding and anti-bonding MOs. The six electrons (3d$65$ 4s$^1$) of the Cr atom are accommodated in the bonding MOs and there is a large HOMO-LUMO gap which results in the stability of the endohedral Cr@B$_{20}$ cage cluster. 

For the Mo@B$_{22}$ cage, the $\pi$-bonded MOs for the bare B$_{22}$ cage and also the MOs for Mo@B$_{22}$ along with the corresponding eigenvalues are shown in Fig. \ref{fig:B22-MO} as obtained from the Gaussian calculation with the PBE0 functional. It is seen that the stability of the Mo encapsulated B$_{22}$ cage is governed by the nine $\pi$-bonded MOs with electronic shell closing at 18 valence electrons. For the bare B$_{22}$ cage there are six occupied $\pi$-bonded MOs and three empty $\pi$-bonded MOs (not shown) of $d$ angular momentum character just above the HOMO. The five $\pi$-bonded MOs with {\it d} angular momentum character hybridize strongly with the {\it d} orbitals of the Mo atom forming five bonding MOs and five anti-bonding MOs. The bonding MOs are fully occupied and similar to the case of 18 valence electron stability of the B$_{20}$ cage, the stability of the Mo@B$_{22}$ cluster also arises from 18 valence electrons. The ordering of the MOs can be seen in the figure. Note that the bare B$_{22}$ cage has very small HOMO-LUMO gap but it is very much increased by Mo encapsulation. Further, the stability of the Mo@B$_{24}$ cluster also arises from the occupation of nine $\pi$ bonded MOs corresponding to 18 valence electrons. A similar result will hold for the case of W doping. We have also performed analysis of the MOs of cr@B$_{18}$ bicapped drum structure, and in this case also the stability is associated with 18 $\pi$ bonded valence electrons. On the other hand the stability of the disk-shaped M@B$_{20}$, for M = Mo and W arises from 20 $\pi$ bonded valence electrons.

\begin{figure*}
\centering
\includegraphics[width=0.6\linewidth,]{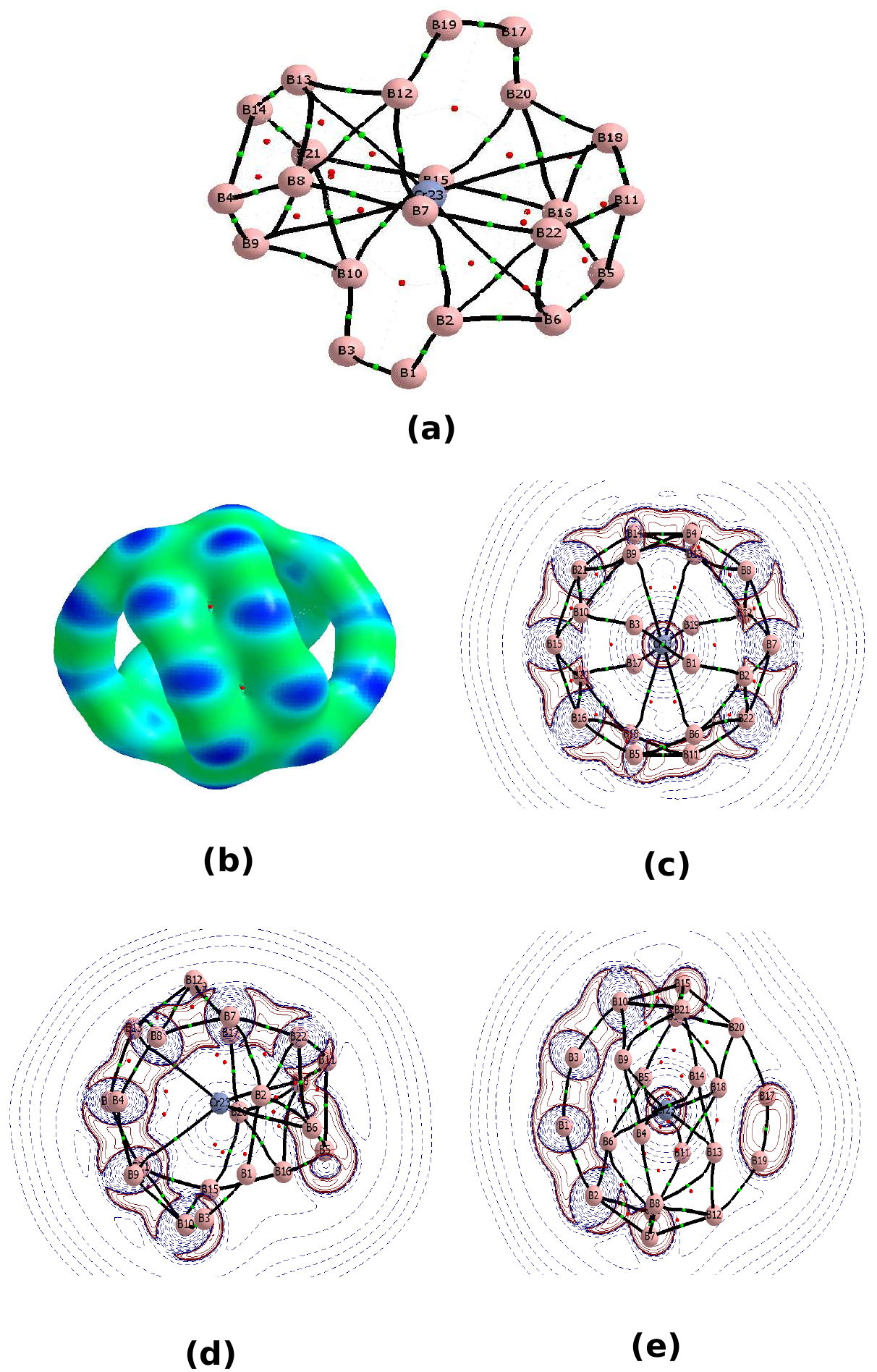}
\caption{(a) BCPs (green dots) and RCPs of BBBB quadrilaterals (red dots) of Cr@B$_{22}$ cluster and (b) electrostatic potential mapped onto a $\rho$(r) = 0.1 e/Bohr$^3$ electron density isosurface. Blue regions indicate negative electrostatic potentials associated with the boron atoms. (c) - (e) Contour plots of the Laplacian in different planes passing through atoms (c) B7, B8, B15, B16, B21, B22, and Cr, (d) B4, B7, B8, B9, B10, B22, and (e) B1, B2, B3, B10, and Cr. The contour values are 0.0, 0.001, 0.002, 0.004, 0.008, 0.02, 0.04, 0.08, 0.2, 0.4, 0.8, 2.0, 4.0, 8.0, 20.0, 40.0, 80.0, 200.0, 400.0, 800.0, -0.001, -0.002, -0.004, -0.008, -0.02, -0.04, -0.08, -0.2, -0.4, -0.8, -2.0, -4.0, -8.0, -20.0, -40.0, -80.0, -200.0, -400.0, -800.0. Pink (grey) balls show B (M) atoms.}
\label{fig:laplacian} 
\end{figure*}

\begin{figure*}
\centering
\includegraphics[width=0.6\linewidth,]{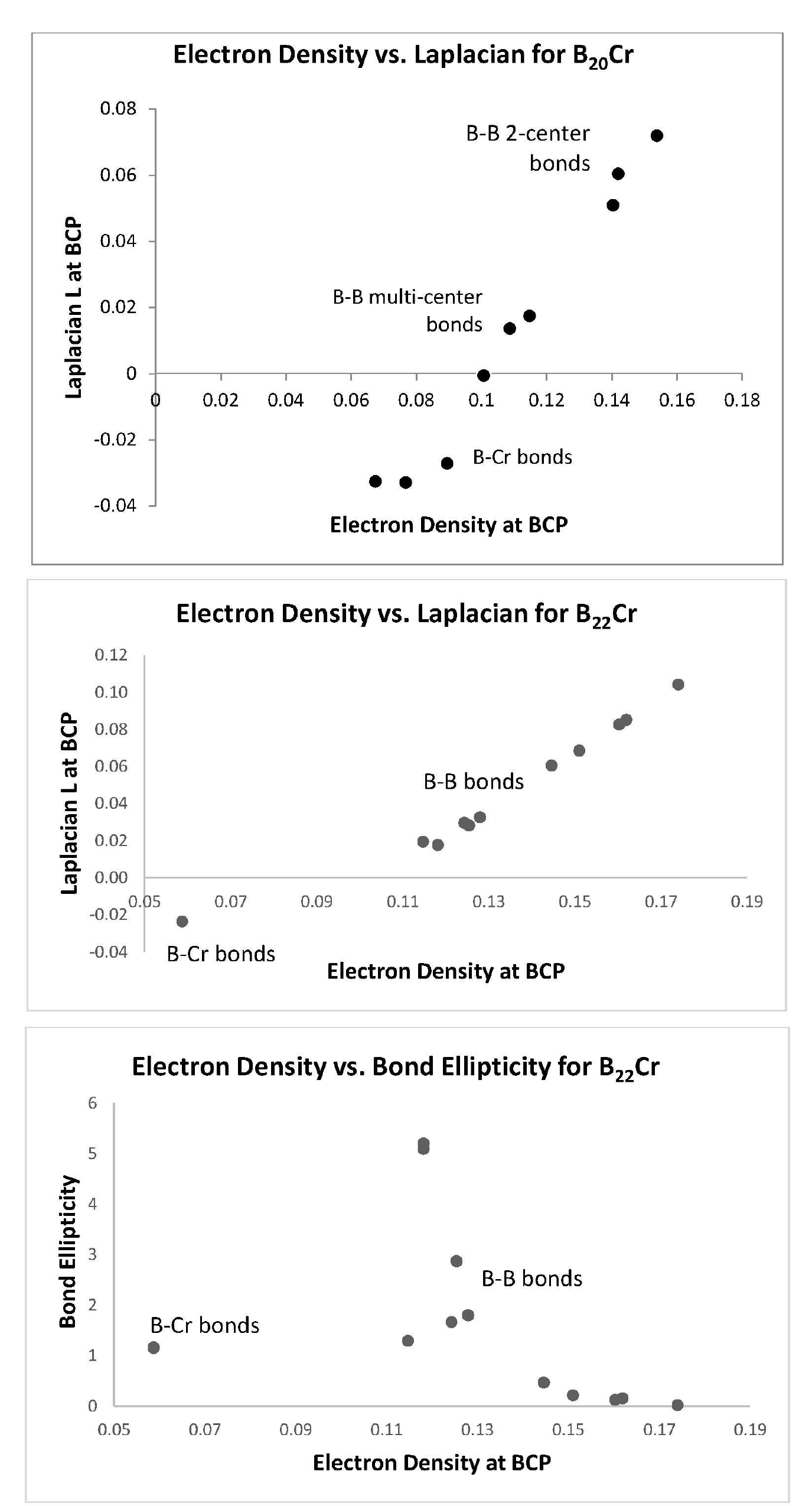}
\caption{(a) Laplacian (in units of e/Bohr$^5$) vs electron density (e/Bohr$^3$), (b) bond ellipticity as a function of the electron density for Cr@B$_{22}$ cluster and (c) Laplacian vs electron density for Cr@B$_{20}$ cluster. }
\label{fig:laplacian1} 
\end{figure*}

We have further performed analysis of the electronic charge density $\rho(\bf{r})$ of Cr@B$_{22}$ as representative of the 22-atom boron cage by calculating contours of the Laplacian of $\rho(\bf{r})$ as well as Laplacian at bond critical points (BCPs) and ring critical points (RCPs). The method of calculation has been discussed earlier.\cite{saha2016} Figure \ref{fig:laplacian} (a) shows the atomic structure with BCPs (green dots) and RCPs (red dots) while (b) shows a symmetrical view of Cr@B$_{22}$ along with the electrostatic potential mapped onto a $\rho(\bf{r}) = 0.1 e/Bohr^3$ electron density isosurface. Blue regions indicate negative electrostatic potential associated with the boron atoms. Of the 22 boron atoms, there are four B atoms (B1, B3, B17, and B19) that have coordination 2 in the middle of the two boron chains, while 18 boron atoms are in two semi-circular bands consisting of BBB triangles and/or BBBB quadrilaterals. The two boron chains appear at the left and right ends of the figure, while one of the semi-circular bands appears in the foreground running diagonally from top left to bottom right (the other is partially occluded at the back, running from top right to bottom left). The two bands and the two chains loop around and encapsulate the M atom in a tetrahedral coordination. Such bands are a recurring motif in many kinds of boron clusters, including drums\cite{saha2016,saha2017} and quasiplanar structures.\cite{B84} The two bands join each other and the two chains at four tetra-coordinate boron atoms. The critical point analysis requires that the Poincare-Hopf relationship: $NumNACP + NumNNACP - NumBCP + NumRCP - NumCCP = 1$ be satisfied, where $NumNACP$ is the number of nuclei (here 23), $NumNNACP$ is the number of non-nuclear attractors (here 0) critical points, $NumBCP$ is the number of BCPs (here 40), $NumRCP$ is the number of RCPs (here 20), and $NumCCP$ is the number of cage critical points (here 2). In Figs. \ref{fig:laplacian} (c)-(e) red dots indicate the locations of the RCPs of the BBBB quadrilaterals. Of the 36 B-B BCPs (green dots), there are 6 that represent covalent bonds in the two boron chains. Figure \ref{fig:laplacian}(d) shows a contour plot of the Laplacian distribution in a plane passing approximately through the rim of one of these bands (Cr@B$_{22}$ with B10-B9-B4-B8-B7-B22). 12 BCPs correspond to bonds at the bases of BBB triangles in the semi-circular bands, and 10 corresponding to bent bonds spanning the width of one or the other of the two bands. Figure \ref{fig:laplacian} (e) (Cr@B$_{22}$ with B2-B1-B3-B10-Cr23) shows a contour plot of the Laplacian distribution in a plane passing approximately through one of the boron chains and the M atom. 8 BCPs represent covalent bonds along rims of the semi-circular bands. The properties of these bonds are shown in Table II and Fig. \ref{fig:laplacian1}. Large values of the charge density and large positive Laplacian with low values of bond ellipticity correspond to covalent bonds while high bond ellipticities, low positive values of the Laplacian L($\bf r$) and delocalization index (off-diagonal localization delocalization matrix (LDM) elements) are characteristic of multi-center bonding in the BBB triangles. On the other hand negative values of Laplacian correspond to the M-Cr bonds. Silimar results have been obtained for Cr@B$_{20}$ cage and shown in Fig. \ref{fig:laplacian1}(c).

\begin{table}
\begin{center}
\caption{Different types of bonds, the number of bonds of each type, charge density ($\rho$) in units of e/Bohr$^3$ at the BCP, bond ellipticity, Laplacian (L) in units of e/Bohr$^5$ at the BCPs, and the delocalization index. Small value of ellipticity and large positive value of Laplacian indicates strong 2c covalent bonds while large values of ellipticity  and small values of Laplacian indicates delocalization of charge (multi-center bonding). The numbering of boron atoms is given in Fig. (\ref{fig:laplacian}).}

\begin{adjustbox}{width=\textwidth}
\begin{tabular}{|l|ccccc|}
\hline
\small{Bond Type}&\small{No. of Bonds}& \small{$\rho$} &\small{Bond Ellipticity} & \small{L}	&\small{Delocalization Index}\\
\hline
\small{B1-B3, B17-B19                 } &2      &0.174 &0.02 &0.104 &1.183\\
\small{B1-B2, B3-B10, B12-B19, B17-B20} &4      &0.162 &0.15 &0.085 &1.023\\
\small{B4-B8, B5-B16, B11-B22, B14-B21} &4     	&0.160 &0.12 &0.083 &1.009 \\
\small{B4-B9, B5-B6, B11-B18, B13-B14 } &4      &0.151 &0.22 &0.069 &0.959 \\
\small{B7-B8, B7-B22, B15-B16, B15-B21} &4	&0.145 &0.47 &0.061 &0.886\\
\small{B2-B6, B9-B10, B12-B13, B18-B20} &4      &0.128 &1.80 &0.033 &0.686\\
\small{B2-B22, B8-B12, B10-B21, B16-B20}&4	&0.125 &2.87 &0.028 &0.555\\
\small{B2-B7, B7-B12, B10-B15, B15-B20} &4	&0.124 &1.66 &0.030 &0.747\\
\small{B9-B21, B6-22, B16-B18, B8-B13 } &4      &0.118 &5.10-5.20 &0.018 &0.555\\
\small{B5-B11, B4-B14                 } &2      &0.115 &1.295     &0.019 &0.572\\
\hline
\end{tabular}
\end{adjustbox}
\end{center}
\end{table}

We also performed calculations on cationic and anionic clusters. For these calculations we have considered the lowest energy and some low-lying isomers of neutral clusters. The cationic and anionic clusters are calculated using Gaussian09 program. Note that for the neutral and charged clusters the calculations using Gaussian09 program give almost the same energy order of the low-lying isomers as with the VASP calculations, but we find that in some cases the lowest energy neutral isomer does not have the lowest energy when charged. As an example the cation of W@B$_{22}$ has a double bicapped drum structure, while the anion of the lowest energy neutral isomer continues to have the lowest energy. Also a cage structure of anions of M@B$_{20}$ (M = Mo and W) has the lowest energy. More details are given in Supplementary Information. We also performed calculations on isoelectronic neutral and anionic clusters of B$_{18}$, B$_{20}$, and B$_{22}$ with the doping of V, Nb, and Ta atoms using Gaussian 09 code with PBE0 functional. Broadly the trends are similar as obtained for the cases of Cr, Mo, and W doping. For VB$_{18}$ anion we obtained a bicapped drum to be most favorable as for isoelectronic Cr doping, but neutral cluster favors a drum structure over the bicapped drum isomer by 0.27 eV. On the other hand for Nb and Ta doping, the drum isomer is lowest in energy for both neutral and anion similar to the case of isoelectronic Mo and W, respectively. In the case of VB$_{20}$ anion and neutral, a bicapped drum is 0.375 eV and 1.281 eV, respectively, lower in energy than a cage structure which is the most favorable for the neutral Cr case. However, for Nb and Ta doped B$_{20}$ anions a bicapped drum is 3.084 eV and 3.065 eV, respectively, lower in energy than a cage isomer similar to the case of Mo and W doping. Also in both cases the neutral NbB$_{22}$ and TaB$_{22}$ clusters also favor the bicapped isomer over the cage isomer by 2.181 and 1.395 eV, respectively. Interestingly for the B$_{24}$ cage all the three metal atoms V, Nb, and Ta stabilize it without much distortion for both neutral and anions and the electric dipole moment in all the cases is very close to zero. This is in contrast to Cr in which case the M atom drifts away from the center and does not interact with all the B atoms properly.

\begin{figure*}
\centering
\includegraphics[width=1.0\linewidth,]{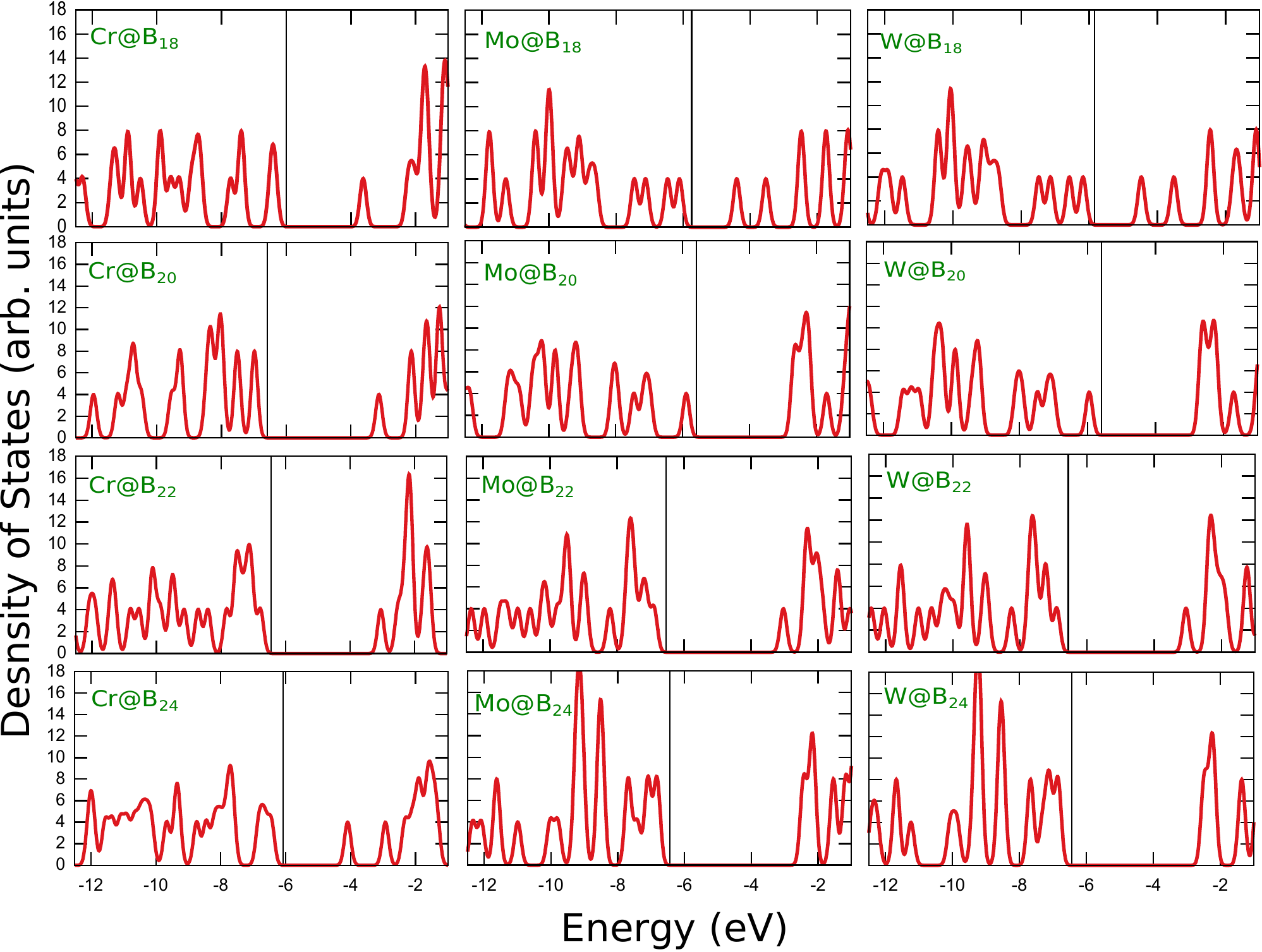}
\caption{Calculated density of states (DOS) for neutral Cr, Mo, and W encapsulated B$_{20}$-B$_{24}$ clusters. The vertical line shows the HOMO.}
\label{fig:DOS-neutral} 
\end{figure*}

The density of states (DOS) for the neutral and anionic clusters provide further information on the nature of the electronic states. The DOS for the anionic clusters (shown in the Supplementary Information) will be useful for comparison with the results of photo-electron spectroscopy experiments that may become available in the future. The DOSs for the neutral clusters with M = Cr, Mo, and W are shown in Figure \ref{fig:DOS-neutral}. Our results show that all the cages have a large HOMO-LUMO gap as obtained from the PBE0 calculations, indicating their very good chemical stability. As seen from the MOs, all the energy levels near the HOMO (shown by the vertical line) arise from hybridization of the boron cage orbitals with the {\it d} orbitals of the M atom. For Cr@B$_{20}$ very localized peaks can be seen below the HOMO, indicating the degeneracy arising from the symmetry of the structure. In going from Cr@B$_{20}$ to Cr@B$_{22}$ there is a spread in the distribution of the electronic energy levels. For Mo@B$_{20}$ and W@B$_{20}$, a capped drum-like structure is lower in energy and the electronic states show more localized peaks compared to the Mo@B$_{18}$ and W@B$_{18}$ cases which lack symmetry. For M@B$_{22}$ and M@B$_{24}$ some sharp peaks can be noticed. The overall nature of the DOS for Mo and W encapsulation remains similar.

\subsection{Vibrational spectra}

\begin{figure*}
\centering
\includegraphics[width=1.0\linewidth]{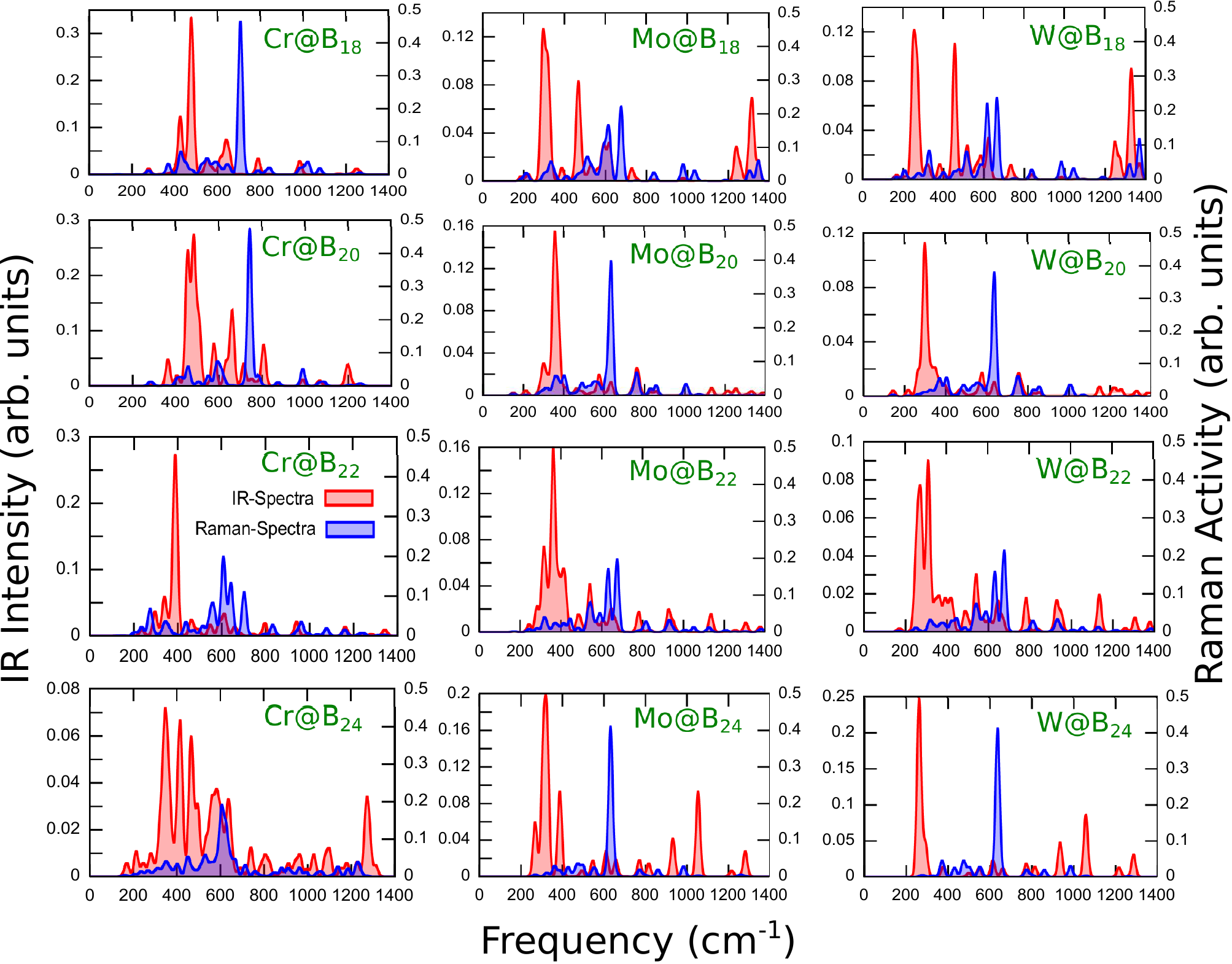}
\caption{Calculated IR and Raman spectra for neutral clusters. }
\label{fig:IR+Raman-neutral} 
\end{figure*}

We have calculated the vibrational modes for the lowest energy neutral isomers and their cations for all the cases using PBE0 functional in Gaussian09 program. Calculations were also done using B3PW91 hybrid functional in Gaussian09 code for some cases. The general features of the spectra using the two functionals are similar, as can be seen from the data in Tables S1-S4 in Supplementary Information, though there are some deviations in intensities and frequencies. Figure \ref{fig:IR+Raman-neutral} shows the calculated IR and Raman spectra for Cr, Mo, and W doped neutral clusters obtained using the PBE0 functional. In almost all the cases we find no imaginary frequency, indicating that the cage structures are dynamically stable. The IR and Raman spectra for the cationic clusters are given in Fig. S9 in Supplementary Information using the PBE0 functional. Tables S1 and S2 in Supplementary Information give the IR intensities and Raman activities for the neutral and cationic clusters, respectively, using B3PW91 functional. We have also given the bond distances in the neutral case. The results using PBE0 functional are given in Tables S3 and S4. Considering the IR active modes for the Cr@$B_{18}$ cluster, we find a strong peak at 424 cm$^{-1}$ corresponding to the breathing mode. In this case the Cr atom vibrates along the axis of the drum, which involves breathing of the two B$_8$ rings. The most dominant modes occur at 473 cm$^{-1}$ and 482 cm$^{-1}$ corresponding to the vibration of the Cr atom in two perpendicular directions in the plane of the B$_{16}$ disk. In one case the Cr atom vibrates parallel to the B-B bond of the two capped boron atoms. 

For the drum-shaped structure of Mo@B$_{18}$ there are strong bending/breathing modes in the range of 291-320 cm$^{-1}$ in the IR spectrum involving M atom and the boron rings. The strong peak at 466 cm$^{-1}$ corresponds to the stretching mode of the boron ring. Another stretching mode occurs with high intensity at 1317 cm$^{-1}$. In the case of W@B$_{18}$, a similar behavior of the modes has been obtained. The strong modes at 250 cm$^{-1}$ and 269 cm$^{-1}$ for W@B$_{18}$ involve swinging of the W atom in two perpendicular directions in the plane of the boron disk, and therefore the frequencies are reduced compared to the Mo case. The mode at 456 cm$^{-1}$ involves bending of the disk without the movement of the W atom, and so it is less affected compared with the Mo case. On the other hand the frequency of the stretching mode at 1332 cm$^{-1}$ has increased compared to the Mo case, which suggests stronger bonding between the boron atoms. The Raman spectrum of Cr@B$_{18}$ has a very high intensity peak at 707 cm$^{-1}$ corresponding to the symmetric breathing mode of the capped drum boron structure. For Mo@B$_{18}$ the high intensity Raman active peak at 676 cm$^{-1}$ and for W@B$_{18}$ the high intensity peaks at 616 cm$^{-1}$ and 665 cm$^{-1}$ corrpdfond to the breathing modes of the B$_{18}$ ring. These vibrations involve purely boron atoms and not the M atom. 
 
For the IR spectrum of Mo@B$_{20}$ the strong modes occur at 353 cm$^{-1}$ and 357 cm$^{-1}$ and correspond to the swing of the Mo atom in the direction perpendicular to the capping boron dimer and parallel to the dimer, respectively. There is also associated bending of the boron rings. The mode at 371 cm$^{-1}$ is also a swing mode of the Mo atom along the boron dimer. For W@B$_{20}$ the IR spectrum is similar with two strong modes with frequencies 294 cm$^{-1}$ and 303 cm$^{-1}$, with the swing of W atom in the direction perpendicular and along the capping boron dimer, respectively. There is an imaginary frequency for MoB$_{20}$ at -122 cm$^{-1}$ and for W@B$_{20}$ at -100 cm$^{-1}$ but in both the cases the intensity is very small. The Raman spectrum of  Mo@B$_{20}$ has two strong modes at 633 cm$^{-1}$ and 762 cm$^{-1}$ while for W@B$_{20}$ there is one strong mode at 636 cm$^{-1}$. These correspond to the breathing mode of the outer capped drum structure.

For Cr@B$_{20}$, the IR spectrum has several high intensity peaks with the most dominant ones at 454 cm$^{-1}$, 455 cm$^{-1}$, 482 cm$^{-1}$, 484 cm$^{-1}$, and 508 cm$^{-1}$ corresponding to the swing modes of the Cr atom while the peaks at 577 cm$^{-1}$ and 662 cm$^{-1}$ correspond to bending modes, and the peak at 807 cm$^{-1}$ represents stretching and breathing mode. In general the IR active peaks involve stretching and bending vibrations of different B-B and B-Cr bonds. However, the Raman spectrum has only one intense peak at 742 cm$^{-1}$ which corresponds to the breathing mode of the cage. For Cr@B$_{20}$ the Raman active peaks come from vibrations involving only boron atoms. For Cr@B$_{22}$, the IR spectrum has dominant peaks at 383 cm$^{-1}$ and 391 cm$^{-1}$, whereas the Raman spectrum has dominant peaks at 608 cm$^{-1}$, 641 cm$^{-1}$, and 702 cm$^{-1}$. Similar spectra are obtained for Mo and W doped B$_{22}$ cages. For Mo@B$_{22}$, the IR modes at 313 cm$^{-1}$ and 362 cm$^{-1}$ are swing modes of the Mo atom with associated motion of the B atoms. The Raman spectrum of Mo@B$_{22}$ has two major peaks at 631 cm$^{-1}$ and 675 cm$^{-1}$ corresponding to breathing mode of the boron cage. For W@B$_{22}$ the dominant IR modes occur at frequencies 259, 275, and 309 cm$^{-1}$, whereas the Raman activity has strong peaks at 630 cm$^{-1}$ and 675 cm$^{-1}$ corresponding to the breathing modes of the cage. In general the dominant IR peaks shift towards lower frequencies compared to Cr encapsulation because of the higher mass of Mo compared to Cr atom. The peaks shift further towards lower frequencies for W encapsulation with IR active modes at 259 cm$^{-1}$ and 275 cm$^{-1}$, compared to 313 and 362 for MO@B$_{22}$ and 383 as well as 391 for Cr@B$_{22}$. 
 
A similar behavior has been obtained for Mo and W encapsulated B$_{24}$ cages. For Mo@B$_{24}$, the IR active dominant peaks are at 325 cm$^{-1}$ and 327 cm$^{-1}$, whereas the Raman spectrum has a strong peak at 632 cm$^{-1}$ corresponding to breathing of the cage. For W@B$_{24}$ the strong IR modes are at frequencies 258 cm$^{-1}$, 263 cm$^{-1}$, 265 cm$^{-1}$, and 1056 cm$^{-1}$, whereas there is one strong Raman peak at 636 cm$^{-1}$ similar to the Mo@B$_{24}$ cage. In general the highest Raman intensity peak corresponds to the breathing mode, whereas the IR active and other Raman active peaks involve different stretching and bending modes. The IR and Raman spectra have more localized and high intensity peaks for the symmetric cases compared to the broader spectrum for clusters with low symmetry.

We also calculated the vibrational spectra for the cationic clusters (see Supplementary Information); Figure S9 in Supplementary Information shows the IR and Raman spectra. For Cr doped cases the IR and Raman spectra differ more significantly for neutral and cationic clusters as compared to those with Mo and W doping. For Cr@B$_{18}^+$ there is one dominant IR peak at 479 cm$^{-1}$ corresponding to swinging mode (perpendicular to the capping boron dimer) of the Cr atom, while the peak at 544 cm $^{-1}$ corresponds to swinging along the boron dimer and the peak at 638 cm$^{-1}$ corresponds to the motion of the Cr atom along z direction. There is only one strong Raman mode at 717 cm$^{-1}$ corresponding to breathing of the boron cage. In the case of Mo@B$_{18}^+$ the IR spectrum has dominant peaks at 312 cm$^{-1}$, 313 cm$^{-1}$ (swing of the M atom), 481 cm$^{-1}$ (swing of the cage), and 730 cm$^{-1}$ (stretching and bending of the B-B bonds), while the Raman spectrum has a strong peak at 676 cm$^{-1}$ corresponding to the breathing mode of the boron cage. In the case of W@B$_{18}$ cation, the strong IR modes are at 268 cm$^{-1}$ and 269 cm$^{-1}$ corresponding to swinging of the W atom, while the mode at 462 cm$^{-1}$ corresponds to the swing of the cage. The Raman spectrum has one strong peak at 667 cm$^{-1}$ corresponding to the breathing mode.  

For Cr@B$_{20}$$^+$ the dominant IR modes are at 402 cm$^{-1}$ corresponding to the swing of the M atom, 748 cm$^{-1}$ (bending of the boron network), 802 cm$^{-1}$ (swing of the cage), 960 cm$^{-1}$ (stretching of the B-B bonds), and 1208 cm$^{-1}$ (stretching of the B-B bonds). The Raman spectrum has strong peaks at 164 cm$^{-1}$ (swinging of the M atom) which is also IR active, and scissor mode at 960 cm$^{-1}$, and stretching/breathing mode at 968 cm$^{-1}$. For Mo@B$_{20}^+$ the dominant IR modes are at 338 cm$^{-1}$, 354 cm$^{-1}$, and 359 cm$^{-1}$ (all three swing modes of the Mo atom), while the Raman spectrum has a strong peak at 650 cm$^{-1}$ corresponding to the breathing of the cage. In the case of W@B$_{20}^+$, the dominant IR modes are at 292 cm$^{-1}$ and 304 cm$^{-1}$ (corresponding to swinging of the W atom in two perpendicular directions) while the dominant Raman peak is  at 651 cm$^{-1}$ corresponding to the breathing mode. 

In the case of Cr@B$_{22}^+$ the main IR peaks occur at 81 cm$^{-1}$, 321 cm$^{-1}$, and 382 cm$^{-1}$, all corresponding to the swinging of the Cr atom, while the Raman spectrum has a strong peak at 693 cm$^{-1}$ corresponding to the breathing of the cage. There is a imaginary frequency at 187 cm$^{-1}$ but the intensity is very small. For Mo@B$_{22}^+$, the main IR peaks are at 326 cm$^{-1}$ and 369 cm$^{-1}$ corresponding to swinging modes of the Mo atom, while the Raman spectrum has two dominant peaks at 615 cm$^{-1}$ and 677 cm$^{-1}$ corresponding to breathing modes. There is an imaginary frequency at 91 cm$^{-1}$ but again the intensity is low. For W@B$_{22}$ the main IR peaks are at 258 cm$^{-1}$, 297 cm$^{-1}$, and 302 cm$^{-1}$ corresponding to the swinging modes of the W atom, bending mode at 328 cm$^{-1}$, and stretching mode at 1150 cm$^{-1}$. The Raman spectrum has a dominant peak at 678 cm$^{-1}$ corresponding to a breathing mode. For Mo@B$_{24}$ the dominant IR peaks are at 314 cm$^{-1}$, 367 cm$^{-1}$, and 376 cm$^{-1}$ corresponding to swinging modes of the Mo atom, while the dominant Raman peak is at 636 cm$^{-1}$ corresponding to the breathing mode. For W@B$_{24}^+$, the IR modes are strong at 253 cm$^{-1}$, 267 cm$^{-1}$, and 297 cm$^{-1}$ corresponding to swinging modes, while the peak at 1035 cm$^{-1}$ is a stretching mode. The Raman spectrum has a strong peak at 638 cm$^{-1}$ corresponding to the breathing mode. In general for Mo@B$_{22}$$^+$, Mo@B$_{24}$$^+$, W@B$_{22}$$^+$ and W@B$_{24}$$^+$ the positions of the dominant peaks occur at similar frequencies as for the corresponding neutral cases.

\section{CONCLUSIONS}
In summary, we have performed a systematic study on M = Cr, Mo, and W doped boron clusters in the size range of 18 to 24 boron atoms. Our results show that by M encapsulation, it is possible to have fullerene-like cage structures of boron with about 20 atoms in contrast to dominantly planar or quasi-planar or tubular structures of pure boron clusters for less than 40 atoms. Our results suggest that doping of Cr is suitable to produce symmetric small cage clusters of boron with B$_{20}$, whereas B$_{22}$ is the smallest symmetric cage for Mo and W encapsulation, which is magic and is likely to be produced in high abundance in experiments. A symmetric B$_{24}$ cage is also formed with Mo and W encapsulation, as also predicted recently, but the variation in the BE of the clusters suggests that the B$_{22}$ cage has the optimal size for Mo and W encapsulation. There is a large gain in energy when a Mo or W atom is encapsulated in the cage, which also supports the strong stability of the doped clusters. We performed an analysis of the bonding nature in these clusters and found that the cage structures are stabilized by strong interaction between the M atom $d$ orbitals and the $\pi$-bonded MOs of the bare boron cage. From this analysis we find that the stability of the Cr@B$_{18}$, Cr@B$_{20}$, M@B$_{22}$, M = Cr, Mo, and W, and M@B$_{24}$ clusters is associated with 18 $\pi$ bonded valence electrons while for M@B$_{18}$ (M = Mo and W) disk-shaped tubular clusters, the stability is associated with 20 $\pi$ bonded valence electrons. In most cases the IR and Raman spectra for the neutral and cationic clusters show that the cages are dynamically stable. These results as well as those of the electronic levels of the anionic clusters will help to compare our predictions with experiments. We have also studied isoelectronic anion and neutral B$_{18}$, B$_{20}$, B$_{22}$, and B$_{24}$ clusters doped with V, Nb, and Ta. In general the structural trends are similar as obtained for Cr, Mo, and W but for B$_{24}$ we obtained a symmetric cage in all cases. We hope that our results will stimulate experimental work on these M doped disk-shaped and fullerene-like structures of boron.

\section{ACKNOWLEDGMENTS}

We gratefully acknowledge the use of the high performance computing facility Magus of the Shiv Nadar University where a part of the calculations have been performed. ABR and VK thankfully acknowledge financial support from International Technology Center - Pacific. ABR acknowledges international travel support (ITS), from SERB, Govt. of India. We thank Prof. Cherif Matta for providing access to AIMLDM software.

\bibliography{references}{}
\bibliographystyle{unsrt}

\end{document}